\documentclass[12pt]{article}

\usepackage{etex}

\usepackage{amssymb,amsthm,amscd,amsbsy,array}
\usepackage{graphicx}    
\usepackage{rotating}    
\usepackage{bm}          
\usepackage{amsmath}
\usepackage{amsfonts}    
\usepackage{amsbsy}
\usepackage{mathrsfs}

\usepackage[backref]{hyperref}	

\usepackage{graphics,graphicx,xcolor}
\usepackage[all]{xy}

\let\ssection=\section
\renewcommand{\section}{\setcounter{equation}{0}\ssection}

\newcommand{\balpha}{\boldsymbol{\alpha}}
\newcommand{\bbeta}{\boldsymbol{\beta}}

\newcommand{\brho}{\boldsymbol{\rho}}

\newcommand{\cE}{{\mathcal{E}}}

\newcommand{\bF}{{\bf F}}

\newcommand{\bg}{{\bf g}}

\newcommand{\rg}{\mathrm{g}}

\newcommand{\cI}{{\mathcal{I}}}

\newcommand{\bp}{{\bf p}}

\newcommand{\bP}{{\bf P}}

\newcommand{\bx}{{\bm{x}}}

\newcommand{\ds}{{\dot{s}}}
\newcommand{\dds}{{\ddot{s}}}

\newcommand{\SO}{\mathrm{SO}}

\newcommand{\dt}{\dot{t}}
\newcommand{\ddt}{\ddot{t}}

\newcommand{\bv}{{\bf v}}
\newcommand{\bV}{{\bf V}}

\newcommand{\bX}{{\bf X}}

\newcommand{\dx}{\dot{x}}
\newcommand{\ddx}{\ddot{x}}

\newcommand{\by}{{\bf y}}

\def\bv{{\bm{v}}}
\def\bp{{\bm{p}}}
\def\bP{{\bm{P}}}
\def\bg{{\bm{g}}}

\def\beq{\begin{equation}}
\def\eeq{\end{equation}}
\def\beqa{\begin{eqnarray}}
\def\eeqa{\end{eqnarray}}
\def\nn{\nonumber}
\def\barray{\left(\begin{array}}
\def\earray{\end{array}\right)}
\def\ort{{\rm o}}

\def\ben{\begin{equation}}
\def\bea{\begin{eqnarray}}
\def\een{\end{equation}}
\def\eea{\end{eqnarray}}
\def\besub{\begin{subequations}}
\def\esub{\end{subequations}}

\def\p{{\partial}}
\def\v0{\mathbf{0}}

\newcommand{\const}{\mathop{\rm const.}\nolimits}
\newcommand{\half }{\frac{1}{2}}

\newcommand{\Rarrow}{{\quad\Rightarrow\quad}}

\newcommand{\medbox}[1]{\fbox{%
\rule[-10pt]{0pt}{25pt}$\;\;\displaystyle{#1}\;\;$}%
}

\usepackage{color} 

\definecolor{blau}{rgb}{0,0,0.75}         

\newenvironment{redtext}{\color{red}}{\ignorespacesafterend}
\newenvironment{bluetext}{\color{blue}}{\ignorespacesafterend}

\newenvironment{magentatext}{\color{magenta}}{\ignorespacesafterend}
\newenvironment{cyantext}{\color{cyan}}{\ignorespacesafterend}
\newcommand{\gb}{\colorbox{green}}

\newcommand{\bblue}{\begin{bluetext}}
\newcommand{\eblue}{\end{bluetext}}
\newcommand{\bred}{\begin{redtext}}
\newcommand{\ered}{\end{redtext}}
\newcommand{\bmagenta}{\begin{magentatext}}
\newcommand{\emagenta}{\end{magentatext}}
\newcommand{\bcyan}{\begin{cyantext}}
\newcommand{\ecyan}{\end{cyantext}}

\def\?{\quad\gb{\fbox{\texttt{?}}}\quad}

\def\aand{{\quad\text{\small and}\quad}}
\def\where{{\quad\text{\small where}\quad}}

\def\ie{{\;\text{\small i.e.}\;}}
\def\ie,{{\;\text{\small i.e.,}\;}}

\newcommand{\BEQ}{\begin{equation}}     
\newcommand{\BEA}{\begin{eqnarray}}
\newcommand{\BD}{\begin{displaymath}}
\newcommand{\EEQ}{\end{equation}}       
\newcommand{\EEA}{\end{eqnarray}}
\newcommand{\ED}{\end{displaymath}}
\newcommand{\D}{{\rm d}}                
\newcommand{\II}{{\rm i}}               
\newcommand{\vep}{\varepsilon}          
\newcommand{\vph}{\varphi}              
\newcommand{\demi}{\frac{1}{2}}         
\newcommand{\wit}[1]{\widetilde{#1}}    

\newcommand{\fns}{\footnotesize}        
\renewcommand{\vec}[1]{\boldsymbol{#1}} 

\begin{document}

\title{\centerline{\mbox{{\sc Schr\"odinger symmetry:} a historical review}}
~\\[6pt]
}

\author{
C. Duval\footnote{ deceased 
}
\\
\textit{Centre de Physique Th\'eorique, CNRS,
Luminy} Case 907\\
F - 13288 Marseille Cedex 9 (France)\footnote{
UMR 6207 du CNRS associ\'ee aux
Universit\'es d'Aix-Marseille I and  II and  Universit\'e du Sud Toulon-Var; Laboratoire  affili\'e \`a la FRUMAM-FR2291.}
\\[12pt]
M. Henkel\footnote{mailto: malte.henkel@univ-lorraine.fr -- ORCID: 0000-0002-5048-7852}
\\
\textit{Laboratoire de Physique et Chimie Th\'eoriques (CNRS UMR 7198),}
\\
Universit\'e de Lorraine Nancy\\ B.P. 70239, F - 54506 Vand{\oe}uvre-l\`es-Nancy Cedex (France)\\
and\\
\textit{Centro de F\'{\i}sica T\'eorica e Computacional,} Universidade de Lisboa \\
Campo Grande, P - 1749-016 Lisboa (Portugal)
\\[12pt]
P.~A.~Horvathy\footnote{mailto: horvathy@univ-tours.fr -- ORCID: 0000-0002-6337-4494}
\\
\textit{Institut Denis Poisson CNRS/UMR 7013 }
\\
Universit\'e de Tours - Universit\'e d'Orl\'eans\\ Parc de Grandmont, 37200, Tours, (France).
\\[12pt]
S. Rouhani \footnote{mailto: Rouhani@ipm.ir -- ORCID: 0000-0003-4738-2627}
\\
\textit{School of Particles and Accelerators}\\
\textit{Institute for Research in Fundamental Sciences (IPM)}\\
 Tehran, P.O. Box 19395-5531 (Iran).
\\[12pt]
P.~M. ~Zhang\footnote{mailto:zhangpm5@mail.sysu.edu.cn -- ORCID: 0000-0002-1737-3845}
\\
School of Physics and Astronomy, Sun Yat-sen University, Zhuhai, (China)
 \\[20pt]
}


\maketitle

\begin{abstract}
This paper reviews the history of
the conformal extension of Galilean symmetry, now called  Schr\"odinger symmetry.
In the physics literature, its discovery is commonly attributed to Jackiw, Niederer and Hagen (1972).
However, Schr\"odinger symmetry has a much older ancestry:
the associated conserved quantities were known to Jacobi in 1842/43 and its Euclidean counterpart was discovered by Sophus Lie in 1881 in his studies of the heat equation.
A convenient way to study Schr\"odinger symmetry is provided by a non-relativistic Kaluza-Klein-type ``Bargmann'' framework, first proposed by Eisenhart (1929),
but then forgotten and re-discovered by Duval {\it et al.} only in 1984.
Representations of Schr\"odinger symmetry differ by the value $z=2$ of the dynamical exponent from the value $z=1$ found in representations of relativistic conformal invariance.
For generic values of $z$, whole families of new algebras exist, which for $z=2/\ell$ include the $\ell$-conformal Galilean algebras.
We also review the non-relativistic limit of conformal algebras and that this limit leads to the $1$-conformal Galilean algebra and not to the Schr\"odinger algebra.
The latter can be recovered in the Bargmann framework through reduction.
A distinctive feature of Galilean and Schr\"odinger symmetries are the Bargmann super-selection rules, algebraically related to a central extension.
An empirical consequence of this was known as ``mass conservation'' already to Lavoisier.
As an illustration of these concepts, some applications to physical ageing in simple model systems are reviewed.
\end{abstract}

\thispagestyle{empty}
\baselineskip=16pt


\noindent
\textbf{Keywords}: Schr\"odinger symmetry, conformal Galilei algebras, conformal Galilei groups, Bargmann super-selection rules.

\newpage

\tableofcontents

\newpage
\section{Introduction}

Symmetry \cite{Weyl1955} is a central concept in almost all theories of physical systems and their myriad applications.
Symmetries arise either as internal symmetries or else as dynamical symmetries of time and space.
Here, we are interested in the second class and notably in {\em conformal invariance}. Whenever it occurs, conformal invariance
plays a crucial r\^ole in various theories and application, both for its mathematical and physical aspects.
A necessary condition for conformal invariance is scale-invariance,
and this requirement sharply distinguishes scale-invariant systems from those which do not have this property.
Remarkably, scale-invariance does hold in certain conditions which happen to be of importance.
Scale-invariance coupled to the usual symmetries of time and space, the conservation of energy-momentum and unitarity
`normally' leads to the emergence of the conformal group.\footnote{For notable exceptions, see e.g.
\cite{MillerdeBell1993,Riva2005,Gimenez2023}. A systematic discussion on when scale-invariance does imply conformal invariance is given in \cite{Polchinski1988,DorigoniRychkov}.}
In high-energy physics, scale-invariance is significant in certain situations,
for example in deep-inelastic scattering \cite{Bjoerken1969,AltarelliParisi} or more generally at the points of symmetry breaking,
when it becomes exact and (for sufficiently local theories) gives rise to conformal field-theory (CFT) \cite{Boulware1970}.
Conformal field-theory is among the main ingredients of string theory, see \cite{Polchinski1998,Becker2007} and refs. therein.
On the low-energy side, conformal symmetry is useful in the
description of equilibrium critical phenomena \cite{Polyakov1970},
most notably in two spatial dimensions \cite{BPZ84},
and the behaviour of physical systems near criticality \cite{Zamolodchikov1989}.\footnote{References on conformal invariance are legion.
We limit ourselves to point to some well-known reviews/books which include \cite{Ginsparg90,Cardy90} and \cite{diFrancesco97,Henkel99,HenkelKarevski12,Rychkov17,Nakayama2015}. For a historical review see \cite{Kastrup08}.}

Another development deals with the AdS/CFT correspondence \cite{Maldacena,Aharony:2000}  which is a conjectured relationship between two kinds of physical theories~:
it proposes a duality between theories in Anti-de Sitter space (AdS) and conformal field-theories on the boundary of AdS.
This correspondence has been a subject of intense study in theoretical physics, in particular in string theory and quantum gravity.

Conformal invariance has also been explored in field-theories where the equations of motion are invariant under Galilei transformations.
Galilean field-theories display a non-relativistic conformal structure, which can become infinite-dimensional even in space-time dimensions higher than two.
The first known example of this appeared in studies of gravitational physics \cite{Bondi62,Sachs62}.
Furthermore, when matter is coupled with Galilean gauge theories, various sectors emerge in the non-relativistic limit from the parent relativistic theories,
showcasing an infinitely enhanced Galilean conformal invariance
when compared to the relativistic case \cite{Barnich07,Gopa,Martelli10,Barnich13}.
Physically, the associated scale-transformation
\begin{subequations} \label{scal}
\begin{align} \label{scal:conf}
t\mapsto \frac{1}{\delta^2} t \;\; &, \;\; \vec{r}\mapsto \frac{1}{\delta^2} \vec{r}
\end{align}
\end{subequations}
treats time and space on the same basis
and ascribes to them the same scaling dimension.
Algebraically, these transformations give rise to what we shall call loosely {\em conformal Galilean algebra} (CGA).

Special interest will be devoted in this article to another kind of non-relativistic structure,
first identified from the motion of free particles or the heat equation.
Since it also arises in the free Schr\"odinger equation, it is often referred to as {\em Schr\"odinger symmetry}.
The {\em Schr\"odinger group} is defined by the following transformations of time $t\in\mathbb{R}$ and space $\vec{r}\in\mathbb{R}^d$
\beq
t \mapsto t' = \frac{\alpha t +\beta}{\gamma t + \delta} \;\; , \;\;
\vec{r} \mapsto \vec{r}' = \frac{\mathscr{R} \vec{r} + \vec{v} t +\vec{a}}{\gamma t +\delta}
\eeq
with  real parameters $\alpha,\beta,\gamma,\delta$ such that $\alpha\delta-\beta\gamma=1$, the constant $d$-dimensional vectors
$\vec{v}$, $\vec{a}$ and the rotation matrix $\mathscr{R}\in \mbox{\rm SO$(d)$}$. It also contains a Galilean sub-group, but
with a different representation from the one used in the conformal Galilean algebra.
Notably, time and space transform differently under scale-transformations, namely
\addtocounter{equation}{-2}
\begin{subequations}
\addtocounter{equation}{1}  
\begin{align} \label{scal:gal}
t\mapsto \frac{1}{\delta^2} t \;\; &, \;\;
\vec{r}\mapsto \frac{1}{\delta}\vec{r}
\end{align}
\end{subequations}
\addtocounter{equation}{1}
The associated Lie algebra is called the {\em Schr\"odinger algebra}.
The different scalings in (\ref{scal}) are described by a {\em dynamical exponent} $z$: in the relativistic conformal case (\ref{scal:conf})
one has $z=1$, while in the non-relativistic case (\ref{scal:gal}) one has $z=2$.
The anisotropic scaling (\ref{scal:gal}) characteristic for the non-relativistic theory actually follows from the relativistic scaling
(\ref{scal:conf}) in Duval's ``Bargmann'' framework \cite{DHP2}.\footnote{A substantially more involved Newton-Cartan framework \cite{DH-NC}
is needed whenever $z\ne 2$ \cite{Henkel02}, see section~\ref{sec:lsi} for more details.}
The  history of Schr\"odinger symmetry is far from being straightforward.
In this historical review we shall reconsider what appears to us as the foundations and which make it clear that Schr\"odinger symmetry has indeed
emerged well before the birth of Schr\"odinger (1887-1961) and goes back to pre-quantum times. It is
highlighted by the names of Carl Gustav Jacobi \cite{Jacobi} (1804 - 1851),
of Sophus Lie \cite{Lie} (1842-1899) and of Luther Eisenhart (1876 - 1965) \cite{Eisenhart}.

Schr\"odinger symmetry appears to have been
subsequently re-discovered several times, during the first half of the 20$^{\rm th}$ century, in mathematical studies \cite{Appell1892,Goff1927,Ovsiannikov1958}.
Later, it received some attention from the Russian and Ukrainian schools
of mathematical physics, see \cite{Ovsiannikov1980,Fushchich1993,Fushchich1994} and references therein.

In the physical literature, the understanding that the free Schr\"odinger equation has more symmetries than just the Galilei Lie algebra
is consensually attributed  to the papers published in the early 1970s  by Jackiw
\cite{Jackiw:1972cb}, Niederer \cite{Niederer:1972zz}, and Hagen \cite{Hagen:1972pd} (J-N-H), who seem to have been unaware of the earlier work.
These authors found, at almost the same time but independently, that for a free spin-less non-relativistic particle, the operators
\beq
\begin{array}{llcl}
\hat{H} &=&  \frac{m}{2}\dot{\bx}^2  ~~=~~  \frac{1}{2m} \bp^2 &\text{\small Hamiltonian}
\\[6pt]
\hat{D} &=&  \frac{1}{2}\big(\bp\cdot\bx + \bx\cdot\bp\big) - 2t\hat{H} &\text{\small dilatation}
\\[6pt]
\hat{K} &=&  \frac{m}{2}{\bx}^2 -tD - t^2\hat{H} &\text{\small expansions}
\end{array}
\label{o21}
\eeq
are symmetry generators which combine, with those of the Galilei group,
to the Schr\"odinger group.\footnote{The Schr\"odinger symmetry was
extended to spin-$\frac{1}{2}$ particles by L\'evy-Leblond \cite{Levy-Leblond:1967eic,GomisNovell1895,DHP96}, and by Hagen \cite{Hagen:1972pd}.}

Soon after, these initial results were extended  to multi-particle systems \cite{Roman:1972nv,Burdet:1972xd,BuPeSo}.
Potentials were added first in the single-particle oscillator
case by Niederer \cite{Niederer1974}, and then for multi-particle systems by de Alfaro {\it et al.} \cite{Fubini};
notably it was understood that an {\em inverse-square} potential admits non-trivial symmetries.
A list of Schr\"odinger-symmetric systems is given in \cite{Boyer76,Nikitin2001,DHP2} in 3+1 dimensions and in \cite{Zurab,Bihlo2017} in 1+1 dimensions.
Dynamical symmetries of systems of reaction-diffusion equations are given in \cite{ChernihaKing,ChernihaDavydovych}.

Jackiw further extended these results to the field of a Dirac monopole \cite{JackiwMonop} and of a magnetic vortex
\cite{JackiwVortex}\footnote{The $\ort(2,1)$ symmetry noticed by Jackiw is indeed implied by the conformal structure of
non-relativistic space-time \cite{DuvalThesis}.},
and to Chern-Simons vortices \cite{Jackiw:1990mb,DHP1,DHP2,DHP96}.
The Schr\"odinger symmetry in fluid mechanics was studied in \cite{Jackiw:2000mm,Hassaine:2000ti,Horvathy:2009kz}.

A different twist arose when it was understood that Schr\"odinger-invariance,
in the context of dynamical phase-transitions with a naturally realised dynamical dilatation-invariance,
acts as a co-variance principle which determines the form of scaling $n$-point expectation values \cite{Hen94}.
Subsequently, Schr\"odinger-invariance was rediscovered, once more,
in the context of non-relativistic analogues of the Anti-de Sitter/conformal field-theory (AdS/CFT) correspondence
and applied to Fermi gases, see \cite{Bala08,Son08,MinicPleimling08,FuertesMoroz09,LeighHoang09,Moroz11} and references therein.
These examples should be enough to conclude that Schr\"odinger symmetry
has indeed many applications in both high-energy and low-energy physics.

One of the aims of this review is to give a historical overview of the basic ideas of Schr\"odinger-invariance.
Some of them have become text-book knowledge but their deep potential has not always been fully recognised. 
We then outline the relations with more recent developments in the hope that these may stimulate fresh ideas for future research.

We shall therefore begin with historically-oriented summary of the basic ideas to be gleaned from the works
of Jacobi, Lie and Eisenhart, in sections \ref{JacobiSec}, \ref{sec:Lie} and \ref{sec:Eisenhart}.
After briefly recalling Kastrup's contribution in section~\ref{sec:Kastrup}, in section~\ref{JNHBSec} the
important work of Jackiw, Niederer, Hagen and Barut on  Schr\"odinger symmetry of a massive particle is reviewed.
A more recent development involves the construction of Lie algebras of local scale-invariance, for dynamical exponents $z\ne 1,2$, to be taken up in section~\ref{sec:lsi}.
Besides the Schr\"odinger algebra, we are
also led to consider another Lie algebra of space-time transformations,
namely the so-called conformal Galilean algebra. This algebra was indeed
known for a long time on their own right in the theories
of gravitation \cite{Bondi62,Sachs62}.
Starting from an old idea of Barut \cite{Barut}, in section~\ref{CGA}
we discuss how this algebra (and not the Schr\"odinger algebra) can be obtained as a non-relativistic limit of relativistic conformal algebras.
Another important aspect of Galilei- and thus also Schr\"odinger-invariance
is the Bargmann super-selection rule which conserves the non-relativistic mass, in spite of its dynamics being scale-invariant, as described in section~\ref{Lavoisier}.
Section~\ref{ageing} reviews applications of Schr\"odinger symmetry to relaxation processes far from equilibrium and physical ageing.
We conclude in section~\ref{sec:conclusion}. 

\section{Carl Gustav Jacobi}
\label{JacobiSec}

This section summarises some fascinating insights
taken from the lectures delivered by
Jacobi in 1842/43  at the University of K\"onigsberg \cite{Jacobi} on classical mechanics.
Although much of what comes has later become standard textbook material, we shall present it in a self-contained way, and shall insist on its historical value.

The main point of interest is  Jacobi's observation that
for  the particular choice of the scalar potential [he calls it a ``force function''],
\beq
U(r)=\frac{\gamma}{r^2}\,
\label{inversesquare}
\eeq
where $\gamma$ is an arbitrary real constant (which can also vanish),
Newton's equations admit, in addition to what is now called the total energy, $H$,
two additional constants of the motion, namely the classical counterparts of the operators in \eqref{o21},
\besub
\begin{align}
H=\; &\frac{m\dot{\bx}^2}{2}+U(r) = \frac{1}{2m} \bp^2 + U(r)\,,
\label{Uenergy}
\\[2pt]
D=\; & 
\frac{\; \D}{\D t}{\Big(\frac{m}{2}\bx^2\Big)}-2tH\,,
\label{Jacdil}
\\[2pt]
K=\;&\frac{m\bx^2}{2}-tD-t^2H\,.
\label{Jacexp}
\end{align}
\label{Jacdilexp}
\esub
Jacobi derives these conserved quantities from what we would call today \emph{symmetries}
--- a  concept which did not exist by his time in its present form. This line of thought 
anticipates  Noether's approach \cite{Noether} by 75 years. However, Jacobi's observations hardly attracted any attention at that time, and appear to
have been forgotten before they were re-discovered about 180 years later.

\smallskip

Jacobi  starts with $N$ particles with coordinates\footnote{Herein, $a=1,\ldots, N$ labels the particles,
each having coordinates $x_i^a,\, i=1,2,3$.}
$\bx^a=(x_i^a)$. Multiplying Newton's equations by ``virtual displacements'' $\delta\bx^a = (\delta x_i)^a$
he deduces what textbooks nowadays refer to as the `Principle of Virtual Work',
\beq
\sum_{a,i}\left(m_a\frac{\D^2x_i^a}{\D t^2}-F_i^a\right)\delta x_i^a=0\,.
\label{Jacobi1}
\eeq
Then he assumes that the forces $F_i^a$ do not depend on time explicitly but rather derive instead from a
scalar potential $U$ (Jacobi's `force function'),  $F_i^a=-\frac{\p U}{\p x_i^a}$.
Then (\ref{Jacobi1}) can be re-written  [eqn (2.) of Lect. 2] as,
\beq
\medbox{
\sum_{a,i}m_a\frac{\D^2x_i^a}{\D t^2}\delta x_i^a
=-
\sum_{a,i}\frac{\p U}{\p x_i^a}\delta x_i^a
=-\delta U\,.}
\label{Jac2.2}
\eeq
After this preparation, Jacobi deduces various properties associated with clever choices of the virtual displacements.
 \goodbreak

\medskip
$\bullet$ \textit{Motion of the centre of mass} (CoM).
Jacobi, in his Lecture 3, assumes that the ``force function'' only depends
on the relative positions, $U=U(\bx^a-\bx^b)$, and
derives what he calls ``Das Princip\footnote{`Prinzip' in present German orthography.}  der Erhaltung der Bewegung des Schwerpunkts'' [Principle of the
conservation of the motion of the centre of mass].
To this end, he chooses virtual displacements which correspond to shifting all positions by the same amount,
\beq
\delta \bx^a = \delta \bx,
\quad a =  1, \dots, N \,,
\label{xtranslation}
\eeq
which plainly leaves the coordinate-differences $\bx^a-\bx^b$ invariant. Then \eqref{Jac2.2} becomes
\beq
\left(\sum_{a}m_a\frac{\D^2\bx^a}{\D t^2}\right)\cdot\delta\bx
=-\left(\sum_{a}\frac{\p U}{\p \bx^a}\right)\cdot\delta \bx =0\,,
\label{Jac3.1}
\eeq
whose right-hand-side vanishes due to the antisymmetry $a \leftrightarrow b$.
But  $\delta \bx$ is arbitrary and therefore,
$ 
\sum_{a}m_a\frac{\D^2\bx^a}{\D t^2}=0\,.
$ 
It follows that the \emph{centre of mass} (CoM),  defined as
\beq
\bX=\sum_a\frac{m_a\bx^a}{M} \where M=\sum_am_a \,
\label{CoM}
\eeq
moves  freely [eq. (2.) Lect.3 in \cite{Jacobi}],
 \beq
 \medbox{
\frac{\D ^2\bX}{\D t^2}= 0
\Rarrow
\bX(t) = \balpha + \bbeta{t}\,,
\qquad \balpha, \bbeta=\const
}
\label{freeCoMmot}
\eeq
to which Jacobi refers to as ``Das Prinzip  der Erhaltung der Bewegung des Schwerpunkts'' [Principle of CoM conservation].
\footnote{If the forces do not come from  potentials -- today we would say that the system is not conservative
-- then the right-hand-side of \eqref{Jac3.1} is replaced by
$
M {\D^2\bX}/{\D t^2} = \sum_a \bF^a
$
[\cite{Jacobi} eq.(3.4)].}

Jacobi {not} does spell out in detail what has become, after Noether,  the standard consequence drawn from invariance with respect to  translations,
\eqref{xtranslation},   --- namely \emph{momentum conservation}.
However this  would  follow at once by setting (no sum over $a$)
\beq
\bp^a = m_a\frac{\D\bx^a}{\D t}
\aand
\bP =\sum_{a}\bp^a =
M\,\frac{\D\bX}{\D t}
\label{momenta}
\eeq
and rewriting  the previous formul{\ae} as,
\beq
\frac{\ \D}{\D t}\!\left(\sum_{a}\bp^a\right)=
\frac{\D\bP}{\D t} =0\,.
\label{momcons}
\eeq
Eq.  \eqref{momcons} goes beyond the mere conservation of the total linear momentum:
the latter depends only on the CoM dynamics but not on the internal motions \footnote{Comparing with recent work on Carrollian systems  \cite{Duval:2014uoa}
shows that while  a single Carroll particle cannot  move \cite{Leblond1965} , however systems composed of several Carroll particles can have non-trivial internal motion
\cite{multiCarroll}. The clue is Carrollian boost symmetry, see section 4.1 of \cite{MZCH}.}.
Jacobi's presentation anticipates  Souriau's \emph{d\'ecomposition barycentrique} into CoM and relative motion \cite{SouriauSSD}.

\medskip
$\bullet$ ``\textit{Theorem of the living force''}.
In his Lecture 4 Jacobi then considers the virtual displacements
\beq
\delta \bx^a =  \frac{\D\bx^a}{\D t}\,\delta{t}
\label{ttransl}
\eeq
for which he deduces from \eqref{Jac2.2}
\beq
\sum_{i,a} m_a\left\{\frac{\D^2x_i^a}{\D t^2}\frac{\D x_i^a}{\D t}\right\}= -\frac{\D U}{\D t}
\nn
\eeq
that he integrates to get
what he calls the  \emph{``Das Prinzip  der Erhaltung der lebendigen Kraft''} [Principle of conservation of the living\ force] defined as
$\sum_{a} m_a\bv_a^2\,
$
where  $\bv_a={\D\bx^a}/{\D t}$,
\beq
\medbox{
\half \sum_{a=1}^{N} m_a{\bv_a^2}
+ U ={\cE}\,,}
\label{energycons}
\eeq
where $\cE$ is a  constant of the motion.
Subtracting eq. \eqref{energycons} for two different moments $\cE$ is eliminated,
showing that the variation of (half of) the living force equals that of the ``force function'' at the end points
--- what we call now the \emph{work} of those forces.

In eq. \eqref{ttransl} we recognise an \emph{infinitesimal time translation};
the ``living force''  is twice the \emph{kinetic energy}, and
$\cE$  is the \emph{total  conserved energy}.
Henceforth we follow the present-day terminology instead of the historical one.

Introducing the  velocity of the CoM and the relative coordinate measured from the latter,
\beq
\dot{\bX} =\bV = \sum_a \frac{m_a \bv_a}{M}
\aand
\brho_a = \bx^a-\bX\,,
\label{coordshift}
\eeq
respectively. Note that $\bx^a - \bx^b=\brho^a-\brho^b$
is independent of the choice of the origin.

After some manipulations which involve also \eqref{freeCoMmot}, we find that \eqref{energycons} can also be presented in a form
\beq
\half \sum_a m_a \left(\dot{\brho}^a\right)^2+U=\widetilde{\cE}
\label{Jac4.8}
\eeq
where $\widetilde{\cE}$ is a redefined constant.
Then the conserved energy  \eqref{energycons} is decomposed into the sum of a CoM and of an internal part,
\beq
\cE= {\cE}^{CoM}+ {\cE}^{int} =
\dfrac{1}{2} M\bV^2
+
\left(\sum_a\dfrac{1}{2}  m_a (\dot \brho_a)^2 + \sum_{a \neq b } U(\brho_a -\brho_b)\right)\,.
\label{energysplit}
\eeq
This decomposition corresponds to that of Souriau in \cite{SouriauSSD} section 13 pp. 162-168.
\goodbreak

$\bullet$ \textit{Das Prinzip der Erhaltung der Fl\"achenr\"aume}
[Principle of area conservation].
In his 5th Lecture Jacobi studies just a particular case : he considers a rotation  in the plane with coordinates
$y = r \cos \varphi, \,  z = r\sin \varphi $ around the $x$ axis by an infinitesimal angle $\varphi \to \varphi + \delta\varphi$. The virtual displacement is  thus
\beq
\barray{c}
\delta y^a
\\
\delta z^a
\earray =
\barray{cc}
0&-z^a
\\
y^a &0
\earray \delta\varphi\,.
\label{angrot}
\eeq
Assuming that the potential is radial in these coordinates,
inserting into \eqref{Jac2.2} and integrating to yield,
\beq
\sum_a m_a \left\{y^a\frac{\D z^a}
{\D t}-z^a\frac{\D y^a}{\D t} \right\} = \cI =\const
\label{Jac4.3}
\eeq
which is Kepler's ``Area Law'' \cite{AstronomiaNova} alias the \emph{conservation} of [the $x$-component of] the \emph{angular momentum}.
He mentions but does not fully work out the intricacies  studied, e.g., by Souriau in section~13 pp. 162-168 of his book \cite{SouriauSSD}
under the title ``d\'ecomposition barycentrique'' [barycentric alias CoM decomposition] which goes substantially beyond our historic study and is therefore omitted.

We mention nevertheless a remarkable footnote of Jacobi [on his p.34], in which he notes that the Area Law discussed above does not, strictly speaking,
apply even to the Solar system, because there is no fixed point in the Universe.
However, it remains  valid when the origin of the coordinates is displaced to the CoM   discussed in the next item.

\medskip
$\bullet$ It is hardly surprising that Jacobi does \emph{not} emphasise \emph{Galilei boost symmetry},
and just notices {\sl en passant} the invariance  under the coordinate change
\beq
\bx^a \to
\widetilde{\bx}^a = \balpha + t \bbeta\,,
\label{transboost}
\eeq
where $\balpha,\, \bbeta $ are constant $3$-vectors. Then he argues that this freedom allows us to shift the origin of the coordinate system to the CoM \eqref{CoM}
or conversely and closer to Galilei's spirit, switching to a co-moving frame where the CoM is at rest,
$ 
\widetilde{\bX}= 0\,.
$ 
Nor does he consider Noether-type conserved quantities.
However we note (anachronistically) that  using \eqref{momenta} and \eqref{momcons} we could check directly
that
\beq
\bg = \left(\sum_a \bp^a\right)t - \left(\sum_a m_a \bx^a\right)
= \bP\,t - M\bX
\eeq
is a constant of the motion, $\D\bg/\D t=0$, which depends only on the CoM.

\medskip\goodbreak

$\bullet$ \textit{Conformal extensions}.

Jacobi's  genuinely {\em new} observation {(see p. 21 in his 4$^{\rm th}$ lecture)} which has  long escaped attention  comes from
assuming that the potential is homogeneous of degree~$k$
\beq
U(\lambda \bx)=\lambda^k U(\bx), \quad\lambda>0 \quad
\text{\small which implies}\quad
\sum_{a,i}x_i^a\frac{\p U}{\p x_i^a}=k\,U\,.
\label{homkpot}
\eeq
Then Jacobi suggests first to \emph{dilate  all position coordinates in the same proportions},
\beq
\delta x_i^a=\lambda\, x_i^a\,,\quad\lambda>0 \,.
\label{infxdil}
\eeq
Thus
$
\delta U=\lambda\,k\,U\,,
$
so that \eqref{Jac2.2} requires
\beq
\sum_{a,i}\half{m_a}x_i^a\,\frac{\D^2x_i^a}{\D t^2}=-k\,U\,.
\label{Jacobi4-1bis}
\eeq
\smallskip
Combining with energy conservation \eqref{energycons} allows one to infer,  {(eq. (2.) on p. 22 in his 4$^{\rm th}$ lecture)}
\beq
\medbox{
\frac{\ \D^2}{\D t^2}\left(\half\sum_{a}{m_a}
(\bx^a)^2\right)=-(k+2)U+2\cE\,,
}
\label{Jac4.2}
\eeq
which looks like a virial theorem.

Then Jacobi observes that
for $k=-2$ i.e. for the \emph{inverse-square potential} \eqref{inversesquare} the $U$-term  is switched off;
integrating by $t$  \eqref{Jac4.2} can be rewritten, using the conservation of $\cE$ along the trajectory,
\beq
\frac{\D D}{\D t}=0\,
\where
D =
\sum_{a}
\frac{\;\D}{\D t}\left(\frac{m_a}{2}(\bx^a\big)^2\right)-2t\,\cE\, =\sum_a \vec{p}^a\cdot \bx^a -2t\,\cE\,,
\label{multiD}
\eeq
providing us with an $N$-particle generalisation of the conserved quantity\footnote{140 years later Niederer \cite{Niederer1974}
derived  dynamical symmetries for a particle with a time-dependent potential $V(t,\bx)$.
In particular, if $V(t,\bx)=g(t) |\bx|^{k}$, he found the presence of both dilatation and expansions for the case $k=-2$ and $g(t)$ constant \cite[eq. (3.6)]{Niederer1974}.
For the Newtonian potential with $k=-1$ the dilatation symmetry requires a time-dependent gravitational constant $g(t)$.
But if the Newtonian potential varies as $g(t)\sim t^{-1}$, expansions are a dynamical symmetry \cite[eq. (3.8)]{Niederer1974}.
On the other hand, if the potential has a time-dependence $g(t)\sim t^{-1/2}$, there is a dilatation dynamical symmetry \cite[eq. (3.9)]{Niederer1974}.
The first of those cases was observed independently in
\cite{DGH91},  consistently with an earlier suggestion of Dirac \cite{Dirac1/t}.}
\label{Niedfoot}
\eqref{Jacdil} and we also recall the definition $\vec{p}_a=m_a\dot{\bx}^a$ from (\ref{momenta}).

Then Jacobi finds, after some manipulations which involve also \eqref{freeCoMmot},
the decomposition [Eq. (6.) {on p. 23} in his section 4.]
\beq
\sum_a m_a (\bx^a)^2 =  M\bX^2 + \sum_a m_a (\brho^a)^2
\label{Jac4.6}
\eeq
where $\brho^a$ is the shifted coordinate with respect to the CoM in \eqref{coordshift}. 
Combining with \eqref{energysplit} yields, in conclusion, the decomposition \footnote{The term in the first line of \eqref{Ddecomp} can be rewritten as
$M\frac{\D\bX}{\D t}\cdot\bX = \bP\cdot\bX$.
 },
\beqa
D &=  &\underbrace{\half\frac{\D\big({M\bX^2}\big)}{\D t}
-tM\bV^2}_{\mbox{\rm \fns CoM}}
\nonumber \\[6pt]
&+&
\underbrace{\left\{\sum_a \half m_a \frac{\D\big(\brho^a\big)^2}{\D t}
- t\left(\sum_a m_a (\dot \brho_a)^2 +2 \sum_{a\neq b} U(\brho_a -\brho_b)\right)\right\}
 }_{\mbox{\rm\fns internal}} .\quad
\label{Ddecomp}
\eeqa
For a single particle the  internal terms duly vanish and the one-particle expression  \eqref{Jacdil} is recovered.

\medskip
The next step could be to consider \emph{expansions}, \eqref{Jacexp}. However by recalling that
$E$ and $D$ are both conserved allows us to deduce at once that
\BEQ
\frac{\ \D}{\D t}
 \left(\frac{m\bX^2}{2}-tD-t^2\cE\right)
=
\bP\cdot\bX- D -2t\,\cE
=0
\EEQ
so that the bracketed quantity,
\begin{equation}
K=\frac{m\bX^2}{2}-tD-t^2\cE,
\label{NJacexp}
\end{equation}
is also conserved. 
Inserting here \eqref{Ddecomp} and \eqref{energysplit} would provide us with a (rather complicated and therefore omitted)
CoM + relative motion decomposition of expansions.

\medskip
In conclusion,
$\cE,\,D,\,K$ are precisely the conserved
quantities \eqref{Jacdilexp} one obtains from the Noether theorem applied to the conformal
$\SO(2,1)$  symmetry, \cite{Jackiw:1972cb, Niederer:1972zz,Hagen:1972pd, Burdet:1972xd, Roman:1972nv, BuPeSo},
which will be discussed further in section~\ref{JNHBSec}.

\section{Sophus Lie} \label{sec:Lie}

Lie's ground-breaking paper \cite{Lie} presents itself modestly as a `note' on the integration of the linear partial differential equation
\beq \label{3.1}
R \frac{\partial^2 z}{\partial x^2} + S \frac{\partial^2 z}{\partial x\partial y} + T \frac{\partial^2 z}{\partial y^2} +
P \frac{\partial z}{\partial x} + Q \frac{\partial z}{\partial y} +  Z z =0
\eeq
for a function $z=z(x,y)$ of two independent variables. The function $R=R(x,y)$ is assumed given and similarly for the functions $S,T,P,Q,Z$.
The paper does not contain any references, but sometimes Lie mentions in passing some of his earlier results. The reader is assumed as well to be familiar with the
notation and conventions used. 
Still, the fact that this paper, published in a Norwegian journal, is written in German
should indicate that the author must have hoped for some interest on an international
level.\footnote{Is it admissible to believe that F. Klein (Leipzig) should become impressed by it~?}

Lie \cite{Lie} begins with the ``well-known'' statement that eq.~(\ref{3.1}) can be reduced to one of the two normal forms
\begin{subequations} \label{3.2}
\begin{align}
\frac{\partial^2 z}{\partial x\partial y} +  P \frac{\partial z}{\partial x} + Q \frac{\partial z}{\partial y} +  Z z =0 \label{3.2a} \\
\frac{\partial^2 z}{\partial x^2} +  P \frac{\partial z}{\partial x} + Q \frac{\partial z}{\partial y} +  Z z =0 \label{3.2b}
\end{align}
\end{subequations}
which are then analysed separately (we shall write $z_x=\frac{\partial z}{\partial x}$, $z_y=\frac{\partial z}{\partial y}$ and so on in what follows).
He uses contact transformations $x\mapsto x'=X(x)$, $y \mapsto y'=Y(y)$ and $z\mapsto z'=F(x,y,z,z_x,z_y)$ with a double objective:
(A) reduce the general differential equations (\ref{3.1}) or (\ref{3.2}) to more simple normal forms,
(B) find all contact transformations which map these normal forms onto themselves.
Since this analysis is carried out for infinitesimal transformations, he finds in modern terminology the {\em Lie algebras} of the corresponding groups of transformation.
It follows immediately, and is shown in detail by Lie, that in both eqs.~(\ref{3.2}) one can always set $P=0$. The analysis of the two eqs.~(\ref{3.2}) then proceeds as follows.

The differential equation (\ref{3.2a}) has two distinct characteristics and is then called {\em hyperbolic}.
Lie shows again from his transformation theory that there is a canonical form
$z_{xy} + Q(x-y) z_y + Z(x-y) z=0$ \cite[eq. (10)]{Lie}. Analysing the consequences of the infinitesimal point transformation
\beq
\delta x = \xi(x) \delta\vep \;\; , \;\; \delta y = \eta(y) \delta \vep \;\; , \;\; \delta z = \bigl( z f(x,y) + \vph(x,y) \bigr)\delta\vep
\eeq
where the functions $\xi,\eta,f$ are to be determined (and it is shown that $\vph=0$)
such that (\ref{3.2a}) is transformed onto itself. The calculation is straightforward, if
somewhat lengthy. It turns out that the most general solution reads \cite[eq. (20)]{Lie}
\beq \label{3.4}
\xi(x) = a x^2 + b x + c \;\; , \;\; \eta(y) = \alpha y^2 + \beta y + \gamma
\eeq
with the constants $a,b,c,\alpha,\beta,\gamma$ and $f$ is a linear function.
These are the infinitesimal conformal transformations in $\mathbb{R}^2$. For $Q=Z=0$, eq.~(\ref{3.2a}) reduces to Laplace's equation in the
light-cone coordinates $x$ and $y$. Lie does not write this full set,
he rather considers how to simplify the results through rescalings, and for example
fixes $a=\alpha=1$ along with $b=\beta$ and $c=\gamma$. His final results are given in terms
of the characteristics for the several normal forms considered, simplified with the above rescalings.

The differential equation (\ref{3.2b}) has one characteristic and is then called {\em parabolic}.
First, Lie  shows that with $P=0$, there is the further reduction to the
canonical form $z_{xx} + z_y + Z(x) z=0$. He then analyses the consequences of the infinitesimal point transformation
\beq
\delta x = \xi(x,y) \delta\vep \;\; , \;\; \delta y = \eta(y) \delta \vep \;\; , \;\; \delta z = \bigl( z f(x,y) + \vph(x,y) \bigr)\delta\vep
\eeq
where the functions $\xi,\eta,f$ are to be determined such that (\ref{3.2b}) is transformed onto itself (and it is shown once more that $\vph=0$).
Straightforward calculations lead to a system of differential equations \cite[eq. (27)]{Lie} for
$\xi,\eta,f,Z$. In particular, for $Z=0$ he finds
\beq \label{3.6}
\xi(x,y) = \frac{x}{2} \frac{\D \eta}{\D y} + m y + n \;\; , \;\; \eta(y) =   \alpha y^2 + \beta y + \gamma
\eeq
and
\BEQ
f= \frac{x^2}{8}\frac{\D^2 \eta}{\D y^2} + \frac{m}{2} x - \frac{\alpha}{2} y +\delta
\EEQ
such that the solution depends on the six parameters
$\alpha,\beta,\gamma,m,n,\delta$.\footnote{Notably, the parameters $\alpha,\beta,\gamma$ describe the `conformal' transformations
in the `time' variable $y$, whereas $m,n$ describe spatial translations and Galilei-transformation in the `space' variable $x$.
Finally, $\delta$ describes a phase shift related to the central extension (called `mass' in later sections).}
If $y$ is interpreted as time and $x$ as space, one recognises the infinitesimal transformations
of the (centrally extended) Schr\"odinger group in $1+1$ space-time dimensions. Again, the solutions are written in form of the characteristics.

Lie does recognise the importance when more than one symmetry transformation is possible.
He explicitly mentions the heat equation $z_{xx}+z_{y}=0$ and its extension with a space-dependent potential $z_{xx} + z_{y} +A x^{-2} z=0$.

It appears that both the sets of (ortho-){\em conformal transformations} (\ref{3.4}) as well as of
{\em Schr\"odinger-transformations} (\ref{3.6}) appear among the historically first examples of space-time dynamical
symmetry transformations. In the remainder of Lie's article \cite{Lie},
these symmetries are applied to reduce the solution of either the equations (\ref{3.2}) to quadratures and a connection with
the theory of minimal surfaces is pointed out. The possibility of an extension to more than two variables and/or
to linear differential equations of higher order is mentioned. It must have looked to Lie too immediate to carry out explicitly,
but was added by later generations of mathematicians \cite{Appell1892,Goff1927,Ovsiannikov1958}.

Almost a century later, these calculations were cast into an appealing form by Niederer \cite{Niederer:1972zz,Niederer1973,Niederer1974}.
Motivated by quantum mechanics, the differential equation (\ref{3.1}) is re-phrased as a wave equation
\beq
\mathscr{S} \phi = 0
\eeq
where the form of the `Schr\"odinger operator' $\mathscr{S}$ is read off from (\ref{3.1}). The transformations of the variables $x,y$ is captured in the form of
an infinitesimal generator
\beq
\mathscr{X} = - A(x,y,z)\partial_x - B(x,y,z)\partial_y - C(x,y,z)
\eeq
A solution $\phi=\phi(x,y)$ of (\ref{3.1}) is mapped onto another solution of the same equation, if
\beq \label{3.9}
\bigl[ \mathscr{S}, \mathscr{X} \bigr] = \lambda_{\mathscr{X}} \mathscr{X}
\eeq
with a scalar $\lambda_{\mathscr{X}}=\lambda_{\mathscr{X}}(x,y)$ which depends on the generator $\mathscr{X}$.
In those cases when $\lambda_{\mathscr{X}}=0$, one has a `strong' symmetry of the physical system, where $\mathscr{S}$ plays the r\^ole of a Hamiltonian.
But if $\lambda_{\mathscr{X}}\ne 0$, eq.~(\ref{3.9}) is merely a `weak' symmetry which only applies to solutions of
(\ref{3.1}).\footnote{Niederer does not attempt to identify canonical forms of (\ref{3.1}).
This is probably very much in line with the typical situations met in physics which are described by a specific
{\em representation} of a certain symmetry, rather than the abstract group.}

Solving (\ref{3.9}) leads to a system of differential equations for $A,B,C$. From their solution, the infinitesimal dynamical symmetry transformations are found.
Niederer carried this out for the free Schr\"odinger/free diffusion equation and coined the name
{\em Schr\"odinger group} \cite{Niederer:1972zz}. This treatment very easily allows to
include external potentials (described by a function $Z\ne 0$ in (\ref{3.2b})). For potentials such as $V(x)\sim {x}$ or $V(x)\sim x^2$
the Lie algebra of dynamical symmetries is isomorphic to the one of the free particle \cite{Niederer1973}. For a potential $V(x)\sim x^{-2}$
a true sub-algebra, including dilatations and expansions, but no spatial translations, is found \cite{Niederer1974}.\footnote{Later work extends these considerations
to non-linear generalised heat equations of interest in fluid dynamics
(including e.g. $1D$ Navier-Stokes equation or Burger's equation) \cite{Niederer1978,Martina:2003zr}.
Because of the Cole-Hopf transformation, the projective representations of Schr\"odinger-invariance
relevant for the Burger's equation have additional additive, rather than multiplicative terms \cite{Ivashkevich1997}.}

\renewcommand{\arraystretch}{1.2}
\begin{sidewaystable}
\begin{tabular}{|l|lll|c|l|l|} \hline
group                             & \multicolumn{3}{l|}{coordinate transformations}                                & $\mathscr{S}$ & {\small abbreviations} & R\'ef. \\
\hline
{\small ortho-conformal $(1+1)D$} & $z'=f(z)$    & $\bar{z}'=\bar{z}$                        &
              & $4\partial_z \partial_{\bar{z}}=\partial_t^2+\partial_r^2$ & {\small$z=t+\II r$} & \cite{Lie} \\
                                  & $z'=z$       & $\bar{z}'=\bar{f}(\bar{z})$               &                     & & {\small$\bar{z}=t-\II r$} & \\ \hline
{\small conformal Galilean}       & $t'=b(t)$    & \multicolumn{2}{l|}{$\vec{r}'=\vec{r}\:\dot{b}(t)$}             & & & \cite{Bondi62,Sachs62} \\
                                  & $t'=t$       & $\vec{r}'=\vec{r}+\vec{a}(t)$             &                     & & & \cite{Havas78} \\
                                  & $t'=t$       & $\vec{r}'=\mathscr{R}\vec{r}$             &                     & & & \\ \hline
{\small $\ell$-conformal Galilean} & $t'=b(t)$   & $r'=\bigl( \dot{b}(t)\bigr)^{\ell} r$     &                     & & $\ell \in \frac{1}{2}\mathbb{Z}$ & \cite{Negro97,Henkel02} \\
                                  & $t'=t$       & $\vec{r}'=\vec{r}+\vec{a}(t)$             &                     & & $\ell\ne \frac{1}{2},1$ & \\
                                  & $t'=t$       & $\vec{r}'=\mathscr{R}\vec{r}$             &                     & & & \\ \hline
{\small meta-conformal $1D$}      & $u=b(u)$     & $v'=v$                                    &
              & $\partial_t - \frac{1}{\beta}\partial_{r_{\|}}$ & {\small$u=t$} & \cite{Henkel02}\\
                                  & $u'=u$       & $v'=c(v)$                                 &                     & & {\small $v=t+\beta r_{\|}$} & \\ \hline
{\small meta-conformal $2D$}      & $\tau'=\tau$ & $w'=f(w)$                                 & $\bar{w}'=\bar{w}$  & & {\small$\tau=t$} & \cite{Henkel-meta2} \\
                                  & $\tau'=\tau$ & $w'=w$                                    & $\bar{w}'=\bar{f}(\bar{w})$
              & $\partial_t - \frac{1}{\beta}\partial_{r_{\|}}$ & {\small$w=t+\beta\bigl(r_{\|}+\II r_{\perp}\bigr)$} & \\
                                  & $\tau'=b(\tau)$ & $w'=w$                                 & $\bar{w}'=\bar{w}$
              & & {\small$\bar{w}=t+\beta\bigl(r_{\|}-\II r_{\perp}\bigr)$} & \\[0.12cm] \hline
{\small Schr\"odinger-Virasoro}   & $t'=b(t)$    & \multicolumn{2}{l|}{$\vec{r}'=\vec{r}\,\sqrt{\dot{b}(t)\,}$}    & & & \cite{Jacobi,Lie}\\
                                  & $t'=t$       & $\vec{r}'=\vec{r}+\vec{a}(t)$             &
              & $\partial_t - \frac{1}{2{\cal M}}\Delta_{\vec{r}}$ & & \cite{Hen94}\\
                                  & $t'=t$       & $\vec{r}'=\mathscr{R}\vec{r}$             &                     & & & \\[0.12cm] \hline
{\small meta-Schr\"odinger-Virasoro} & $t'=b(t)$ & $v'=v$ & \multicolumn{1}{l|}{$\vec{r}_{\perp}'=\vec{r}_{\perp}\,\sqrt{\dot{b}(t)\,}$}    & & & \cite{Stoimenov22-metaS}\\
                                  & $t'=t$       & $v'=c(v)$                                 & $\vec{r}_{\perp}'=\vec{r}_{\perp}$
              & $\partial_t - \frac{1}{\beta}\partial_{r_{\|}}- \frac{1}{2{\cal M}}\Delta_{\vec{r}_{\perp}}$ & {\small$v=t+\beta r_{\|}$} & \\
                                  & $t'=t$       & $r_{\|}'=r_{\|}$                          & $\vec{r}_{\perp}'=\vec{r}_{\perp}+\vec{a}(t)$ & & & \\
                                  & $t'=t$       & $r_{\|}'=r_{\|}$                          & $\vec{r}_{\perp}'=\mathscr{R}\vec{r}_{\perp}$ & & & \\ \hline
\end{tabular}
\caption[tab1]{{\small
Examples of infinite-dimensional groups of space-time transformations, defined through abstract coordinate
transformations on time ($t$) and space
coordinates ($\vec{r}\in\mathbb{R}^d$).
A physical bias is parameterised by the constant $\beta\ne 0$ and distinguishes a preferred spatial direction
$r_{\|}\in\mathbb{R}$ from transverse spatial directions $\vec{r}_{\perp}\in\mathbb{R}^{d-1}$.
Time and space transformations are specified in terms of differentiable (vector-valued) functions $f,\bar{f},b,c,\vec{a}$
of their argument, $\dot{b}(t)=\D b(t)/\D t$
and $\mathscr{R}\in\mbox{\sl SO}(d)$ is a rotation matrix.
$\mathscr{S}$ is the invariant Schr\"odinger operator, where $\Delta_{\vec{r}}$ is
the spatial Laplacian.}
\label{tab1}}
\end{sidewaystable}

This approach, based on (\ref{3.9}), can be generalised to different Schr\"odinger operators.
Table~\ref{tab1} lists examples of such symmetry transformations which can be extended, beyond the
finite-dimensional sets considered by Jacobi, Lie and Niederer, to in\-fi\-ni\-te-di\-men\-sio\-nal
Lie algebras (which can be centrally extended).
In particular, the infinitesimal conformal transformations and the Schr\"odinger-transformations
already found by Lie and Jacobi are contained as the maximal fi\-nite-di\-men\-sio\-nal sub-algebras of
the \mbox{(ortho-)}conformal and Schr\"odinger-Virasoro groups, along with the Schr\"odinger operator $\mathscr{S}$
on which they act as dynamical symmetries. Furthermore, given the explicit
space-time transformations, it is easily checked that the full infinite-dimensional ortho- and
meta-conformal transformation groups act as dynamical symmetries of their respective
Schr\"odinger-operator, whereas for the Schr\"odinger-Virasoro and meta-Schr\"odinger-Virasoro groups,
only the maximal fi\-nite-di\-men\-sio\-nal sub-group acts as dynamical symmetry group on $\mathscr{S}$.

\newpage

\section{Eisenhart - Duval - Brinkmann} \label{sec:Eisenhart}

\newcommand{\brg}{\bar{\rg}}

In a path-breaking paper \cite{Eisenhart}, unnoticed until recently by the physics community, Eisenhart has shown that the equations of motion of a quite general
conserva\-ti\-ve, holonomic, dynamical system with $n$ degrees of freedom can be trans\-cribed as the \textit{geodesic equations}
in a certain Lorentzian space-time of dimen\-sion~$n+1$. This discovery which goes back to the late 1920s, used several ingredients borrowed from the then new-born general
relativity theory, and pointed out a remarkable link between the latter and the most classical (non-relativistic) aspects of analytical mechanics.

Eisenhart starts with the mechanical system ruled by the Lagrangian
\begin{equation}
L=\half\,\rg_{\alpha\beta}\,\D x^\alpha\D x^\beta-V
\label{L}
\end{equation}
in space-time configuration space, with coordinates $(x^\alpha)=(x^1,\ldots,x^n,x^{n+1})$, where $t=x^{n+1}$ stands for the absolute time-coordinate.
The quadratic form $(\rg_{\alpha\beta})$ as well as the potential function $V$ are assumed to depend smoothly, albeit arbitrarily, upon $(x^\alpha)$.
It should be stressed that the quadratic form $(\rg_{\alpha\beta})$ is  \textit{not} assumed to be non-degenerate;
however the sub-matrix $(\rg_{ij})$, where $i,j=1,\ldots,n$ represents local\-ly a  \textit{Rieman\-nian metric} on configuration space,
i.e., for each time-slice. Re\-writing the Lagran\-gian~(\ref{L}) as
\beq
L=\half\,\rg_{ij}\,\dx^i\dx^j+\alpha_i\dx^i+\half\varphi-V,
\eeq
where\footnote{Eisenhart singles out $t$ to parameterise the trajectories of the system.} $\alpha_i=\rg_{it}$ and $\varphi=\rg_{tt}$ 
one has the Euler-Lagrange equations
\begin{equation}
\rg_{ij}\ddx^j+\Gamma_{jki}\,\dx^j\dx^k+\big(\partial_t\rg_{ij}+\partial_i\alpha_j-\partial_j\alpha_i\big)\dx^j+\partial_t\alpha_i+\partial_i\big(-\half\varphi+V\big)=0
\label{LagrangeEquations}
\end{equation}
for all $i=1,\ldots,n$, where the $\Gamma_{jki}$ denote the  Christoffel symbols of the metric tensor~$(\rg_{ij})$ on a slice $t=\const$.
The differential equations (\ref{LagrangeEquations}) can be understood as the equations of motion of a mechanical system subject to time-dependent holonomic constraints
(inducing a time-dependent metric $(\rg_{ij})$ for the configuration space), and expressed in a rotating frame
(the $\alpha_i$ representing the coordinates of the ``Coriolis'' vector potential).

The observation of Eisenhart is that the equations of motion~(\ref{LagrangeEquations}) correspond in fact to the geodesic equations of a special Lorentz metric on a
$(n+{2})$-dimensional extended space-time with coordinates $(x^\mu)=(x^1,\ldots,x^n,t,s)$ where $s=x^{n+2}$. Then Greek indices have the values $\mu=1,2,\ldots,n+2$.
The brand-new coordinate $s$ has indeed a mechanical interpretation discovered in \cite{Eisenhart} which is presented below.

The Lorentz metric $\brg=\brg_{\mu\nu}\,\D x^\mu{}\D x^\nu$ introduced in \cite{Eisenhart} reads in fact
\begin{equation}
\brg=\rg_{ij}\,\D x^i\D x^j+2(\alpha_i\,\D x^i+\D s)\D t+A\,\D t^2
\label{brg}
\end{equation}
where the components $\brg_{ij}=\rg_{ij}$ and $\brg_{it}=\alpha_i$ (for $i,j=1,\ldots,n+1)$ depend, along with $\brg_{tt}=A$, on the space-time coordinates
$(x^\alpha)$ only.
The metric (\ref{brg}) turns out to be a \textit{Brinkmann metric} \cite{Brinkmann} characterised by the fact that it possesses a null,
covariantly constant, nowhere vanishing vector field \footnote{This fact seems to have been overlooked by Eisenhart.}
\begin{equation}
\xi=\frac{\partial}{\partial s}.
\label{xi}
\end{equation}
Such a pair $(\rg,\xi)$ has been coined a \textit{Bargmann structure} in \cite{DBKP,DGH91} where it has been devised to desingularize Newton-Cartan structures \cite{Kunzle}.

The equations of the geodesics of the metric (\ref{brg}) readily imply
\begin{equation}
\rg_{ij}\ddx^j+\Gamma_{jki}\,\dx^j\dx^k+\big(\partial_t\rg_{ij}
-\partial_i\alpha_j+\partial_j\alpha_i\big)\dx^j\,\dt+\big(\partial_t\alpha_i-\half\partial_iA\big)\dt^2+\alpha_i\,\ddt=0
\label{ddx}
\end{equation}
for all $i=1,\ldots,n$, together with
$\ddt=0$, and therefore,
$\dt=a$
where $a=\const$.\footnote{Eisenhart assumes $a\neq0$, and never considers the case $a=0$
which would correspond to the ``geodesics'' of a $(n+1)$-dimensional
\textit{Carroll manifold} \cite{Carrollvs} embedded, as a slice $t=\const$,
in the Bargmann space-time extension. Let us furthermore mention that
\BD
\dt=\brg_{\mu\nu}\,\xi^\mu\dx^\nu
\ED
is clearly a constant of the motion since $\xi$, see eq.~(\ref{xi}), is a Killing vector field.
This constant of the motion has been promoted in \cite{DBKP}
to the status of a constant of the whole mechanical system, namely its mass $m$ (one of the Galilei Casimir-invariants \cite{SouriauSSD}).}
Hence the last equation reads as follows
\begin{equation}
\dds+\alpha_j\ddx^j+\half\big(\partial_j\alpha_k+\partial_k\alpha_j-\partial_t\rg_{ij}\big)\dx^j\dx^k+\partial_jA\,\dx^j\,\dt+\half\partial_tA\,\dt^2=0
\label{dds}
\end{equation}
Equations (\ref{ddx}) patently reproduce the dynamical equations~(\ref{LagrangeEquations}) provided one sets
\begin{equation}
A=\varphi-2V
\qquad
\&
\qquad
a=1.
\label{Aa}
\end{equation}
This is Eisenhart's main observation \cite{Eisenhart}.

However, he dispenses with the analysis of Equation (\ref{dds}): instead, he goes on with the interpretation of the co\-ordinate $s$ by considering the first-integral
\begin{equation}
\brg_{\mu\nu}\,\dx^\mu\dx^\nu=c.
\label{c}
\end{equation}
The constant $c$ referred to as the {\em Jacobi invariant} is indeed  a Galilei Casimir-invariant. For a massive mechanical system, it is related to the \textit{internal energy}
$c=-2mE^{int}$ in section~\ref{JacobiSec}, discussed also \cite{SouriauSSD}.
Eisenhart  calls  a geodesic \textit{minimal} when $c=0$ and \textit{non-minimal} otherwise.

With the help of (\ref{brg}), (\ref{Aa}) and (\ref{c}) one finds
\BEQ
\rg_{ij}\dx^i\dx^j+2\alpha_i\dx^i+(\varphi-2V)+2\ds=c.
\EEQ
In view of (\ref{L}), the final result for $s$ is obtained
\begin{equation}
s=-\int{\!L\,\D t}+b-\frac{ct}{2}
\label{s}
\end{equation}
with $b=\const$

Remarkably enough, Eisenhart's approach enabled him to interpret the new variable $s$ as the \textit{classical Hamiltonian action} of the original mechanical system ---
whose familiar expression can be recovered for ``minimal'' (or light-like) geodesics, i.e., those for which $c=0$ as prescribed in \cite{DBKP,EDAHKZ}.
Notice that the latter condition facilitates the emergence of conformal symmetries of the model \cite{DGH91} leading to the geometric defini\-tion of the
Schr\"odinger group on the Lorentzian space-time extension pioneered by Eisenhart. For more details when $c\neq0$ see \cite{EDAHKZ}.

\section{Kastrup} \label{sec:Kastrup}
A different approach to the dynamical symmetries of non-relativistic systems \cite{Kastrup,Kastrup08} deserves to be briefly mentioned.
For a free non-relativistic particle, the kinetic energy $E=\frac{1}{2m} \vec{p}^2$. Hence its (non-constant) velocity is
\beq
\vec{v} = \frac{\partial E}{\partial \vec{p}} = \frac{\vec{p}}{m} = \frac{\D \vec{r}}{\D t}
\eeq
Defining $y^0 := v t$, with $v=|\vec{v}|$, one calls the set of points ${\tt y}=(y^0,\vec{r})$ with the Minkowski metric
${\tt y}\cdot{\tt y}=\bigl( y^0\bigr)^2 - \vec{r}^2$ the `Galilei space'.
Kastrup explicitly gives the infinitesimal space-time transformations in Galilei space and states that their Lie algebra is isomorphic to the conformal algebra
$\mathfrak{so}(2,4)$. The action integral $S = \int \!\D t\: \frac{m}{2} \bigl(\frac{\D\vec{r}}{\D t}\bigr)^2$ is
left invariant under these transformations. From this follow the conservation laws, including those from (\ref{o21}), of the
non-relativistic free particle. This furnishes an example of a representation of the conformal algebra relevant for non-relativistic motion.

\newpage
\section{Jackiw-Niederer-Hagen-Barut}
\label{JNHBSec}

The presence of mass  in the Schr\"odinger equation seems to suggest the absence of scale invariance (as it does in the relativistic case).
This is however not so, as we now explain. The prevalent lore that scale-invariant theories cannot have dimensionful constants does not apply here.
The free Schr\"odinger equation is indeed scale-invariant when time and space are scaled simultaneously with an appropriate dynamical exponent
\cite{Jackiw:1972cb, Niederer:1972zz, Hagen:1972pd, Burdet:1972xd, Roman:1972nv, DBKP,DGH91,Hen94,HenkelPRL97,Henkel02,HenkelUnterberger}.
This section  reviews this (now standard) presentation which follows mostly refs.\cite{Jackiw:1972cb,Niederer:1972zz, Hagen:1972pd}.

The free Schr\"odinger equation, presented in energy-momentum space as
\begin{equation}
\left( \frac{\bp^2}{2m} - E\right) \Phi = 0\,,
\label{FressSchr}
\end{equation}
is clearly scale-invariant under
\begin{equation}
E \to \lambda^2 E\,, \qquad \boldsymbol{p} \to \lambda\, \boldsymbol{p},
\end{equation}
This extends the  well-known Galilei symmetry by scale-invariance.
Then the natural question arises whether this symmetry can further be extended to the conformal group by including also expansions.
We expect that conservation of energy-momentum plus scale-invariance require, just as in the relativistic theory,
conformal invariance, -- and this is indeed the case  also in the non-relativistic limit~\cite{Hagen:1972pd}.
An energy-momentum tensor may be explicitly constructed. It is traceless due to scale-invariance.
Then an inversion operator can be constructed following ref.~\cite{Hagen:1972pd}.

In conclusion, the free Schr\"odinger equation of a massive non-relativistic particle has, beyond the natural Galilean symmetry, two more ``conformal'' symmetries, cf.
\eqref{o21} or \eqref{Jacdilexp},
respectively. Using  Lie algebra language,
the  conserved quantities $D$ and $K$ in \eqref{o21}
close to an ${\rm so}(2,1)$ symmetry algebra with commutation relations,
\beq
\begin{array}{lll}
[\hat{D},\hat{H}] = -2\II\hat{H}, & [\hat{K},\hat{H}] = \II\hat{D},
&[\hat{D},\hat{K}] = 2\II\hat{K}\,.
\end{array}
\label{O21commrel}
\eeq
In ``after-Noether'' spirit, the operators \eqref{o21}
or their classical counterparts \eqref{Jacdilexp}
are associated with the space-time transformations
\BEA
T \to T+\tau\hspace{0.3cm} &,& \bX \to \bx \hspace{1.1cm}\mbox{\rm\small ~~~time translation} \nonumber \\
T \to \lambda^2 T\hspace{0.65cm} &,& \bX \to \lambda \bX \hspace{0.75cm}\mbox{\rm\small ~~~space-time dilatation}  \label{HDK} \\
T \to \frac{T}{1-\kappa T} &,& \bX \to \frac{\bX}{1-\kappa T} \mbox{\rm\small ~~~space-time expansion} \nonumber
\EEA
where $\tau, \lambda, \kappa$ are real parameters and, to distinguish from the Jacobi treatment of section~\ref{JacobiSec},
we introduced new, time and space coordinates $(T,\bX)$ as we shall discuss later.
\goodbreak

Adding dilatations and inversions to the Galilean group spanned by $P_j$ (linear momenta), $J_j$ (angular momenta)
$B_j$ (Galilei boosts) yields a two-parameter extension of the latter, known as the (centre-less) Schr\"odinger group.
Extending the Galilei group by the central element identified as the mass yields the Bargmann ( = centrally extended Galilean)
group \footnote{The mass extension Galilei $\to$ Bargmann is discussed in section~\ref{Lavoisier}.
We just mention that in the plane, there is another ``exotic'' central extension \cite{Martelli10,LLGal,ChernihaHenkel2010}.}.
Combining Bargmann with the \eqref{o21} provides us, at last, with the (extended) Schr\"odinger group {\sl Sch}$(d)$.
The non-vanishing commutation relations of the $\ort(2,1)$
generators \eqref{o21} with those of the Bargmann algebra \cite{Barg54,DBKP,Gomis-Pons} are, in particular (with $j=1,\ldots,d$)
\beq
\begin{array}{lll}
[\hat{D},\hat{J}_j] = 0, & [\hat{D},\hat{P}_j] = - \II\hat{P}_j, &[\hat{D},\hat{B}_j]=\II\hat{B}_j,
\\[6pt]
[\hat{K},\hat{J}_j] = 0, &[\hat{K},\hat{P}_j] = \II\hat{B}_j,
&[\hat{K},\hat{B}_j] =0\,.
\end{array}
\label{Schcommrel}
\eeq

To understand the relation between Jacobi's historic and the presently favoured space-time approaches
we propose to view
the transformations in \eqref{HDK} as a coordinate change.
Restricting  to just a single particle for simplicity ,
\beq
T = \lambda^2 t,\qquad \bX = \lambda\,\bx ,
\label{XxTt}
\eeq
Then Jacobi's conserved quantity  \eqref{multiD}
becomes, with the notation \eqref{energycons} \footnote{$\cE=\cE^{CoM}$ for a one-particle theory.},\hfill\break
$
D= {D}\big(m\bX^2/2\big)/{\D T}-2T {\cE}/{\lambda^2}\,.
$
But  the energy scales also when we switch to the new coordinates
\beq
\cE = \lambda^2 \left(\half m\bigg(\frac{\D\bX}{\D T}\bigg)^2+\frac{\gamma\;}{|\bX|^2}\right)=
\lambda^2  E\,.
\nn 
\eeq
Thus Jacobi's conserved $D$ in \eqref{multiD} is indeed  the same as what we would get from \eqref{XxTt} by Noether,
\beq
D= \frac{\D}{\D T}\left(\half{m}\bX^2\right)-2T E\,.
\label{JNHD}
\eeq
[as anticipated by our notations], whose
conservation follows from $ \D/\D T = \lambda^{-2} \D/\D t$.

These formulae can readily be generalised to multi-particles systems~: it is sufficient to  generalise \eqref{XxTt} to
$\by^a = \lambda\,\bx^a$ and replace $m\bX^2$ by $\sum_a m_a\D (\by^a)^2$ and accordingly for the $T$-derivative.
Then $D$ in \eqref{JNHD} could be further decomposed into CoM and internal parts,
$D = D^{CoM} + D^{int}$,
by using \eqref{energysplit}.

\medskip\goodbreak

Jacobi's clue is the ``virial-type'' formula \eqref{Jac4.2}
he derives from space dilatation alone, \eqref{infxdil} followed by direct integration
-- consistently with Newton's spirit who insists that time is absolute \cite{Newton}.
This is in sharp contrast with today's ``after Einsteinian"  space-time approach, \eqref{HDK}.

The relation of the two approaches is understood  by
recalling the  temporal re-parametrisation scheme in \cite{GaryDark,ZZH},
\beq
t = f(T),
\qquad
\bx = \sqrt{\frac{\D f}{\D T}}\,\bX\,
\label{Gibbscale}
\eeq
For $f(T)= T/\lambda^2$ \eqref{JNHD} is recovered.
In conclusion, the scaling of time in eq.\eqref{XxTt} is
absorbed into the time redefinition, $t\to T$ while the position scales as before.

\goodbreak
Multi-particle systems and their relation to the CoM decomposition, which led Jacobi to study the internal energy and angular momentum and ultimately and
(much later) Souriau's concept of ``classical spin'' \cite{SouriauSpin} go substantially beyond the scopes of our present ``prehistoric'' study.
The interested reader is advised to consult, in particular, section~13 ``D\'ecomposition Barycentrique"  [section~13, pp. 162-168] of \cite{SouriauSSD} for details.

Returning to the algebraic structure, we recall that the suggestion that the correct description of a non-relativistic free quantum particle should involve the
central extension of the Galilei group goes back to In\"on\"u and Wigner~\cite{InonuWigner:1952} and to Bargmann \cite{Barg54}.
In the ray representation of non-relativistic quantum mechanics,
the plane-wave solution of the free Schr\"odinger equation transforms under the action of the  Galilei group (in time and space)
\begin{equation}
t \mapsto t' = t + \beta, \qquad \boldsymbol{x} \mapsto \boldsymbol{x}' = \mathscr{R} \boldsymbol{x} + \boldsymbol{v}+ \boldsymbol{a},
\end{equation}
according to
\begin{equation}
\psi(t,\boldsymbol{x}) \mapsto \exp \bigl[\Lambda(t,\boldsymbol{x})\bigr]\, \psi (t,\mathscr{R} \boldsymbol{x} + \boldsymbol{v}t + \boldsymbol{a}),
\label{rayrep}
\end{equation}
where $\mathscr{R}$ is an SO$(3)$ rotation matrix,  $\boldsymbol{v}$ is the velocity of the inertial frame, and $\boldsymbol{a}$, $\beta$
are shifts in space and time, respectively. These make up the $(\frac{1}{2}d(d+3)+1)$-parameter Galilei group,
which means $10$ group parameters for $d=3$ spatial dimensions.
Because the correct quantum mechanical representation is a ray representation, the phase in eq.\eqref{rayrep} is no longer a  constant but rather reads
\begin{equation}
\Lambda(t,\boldsymbol{x}) = m \vec{v}\cdot \bigl( \mathscr{R}\vec{x}\bigr) + \tfrac{1}{2}m v^2t,
\end{equation}
The presence of the non-relativistic mass $m$ leads to the central extension of the Galilei algebra (which can exist since the non-extended Galilei algebra is not semi-simple).
This means that mass is an operator which commutes with all elements of the algebra~\cite{Jackiw:1972cb,Hagen:1972pd,Niederer:1972zz,Roman:1972nv}, but to the
commutation relations (\ref{Schcommrel}) the non-vanishing commutator
\begin{equation}
\lbrack K_j, P_k \rbrack = \II\, \delta_{jk}M \;\; ; \;\; \mbox{\rm with $j,k=1,\ldots,d$}
\end{equation}
must be added. It is called a {\em central extension}, since $M$ commutes with all generators of the Galilei algebra. It is not possible to transform it away through
a coordinate change \cite{Barg54,UnterbergerRoger2011}.
Furthermore, as explained in more detail in section~\ref{Lavoisier}, this implies that the Bargmann super-selection rules apply~\cite{Barg54}:
\begin{equation}
\sum_j M_j = \sum_j \acute{M}_j,
\end{equation}
where  $M_j$  and $\acute{M}_j$ are the masses before and after an interaction.\footnote{This is the quantum version of Lavoisier's conservation of mass \cite{Lavoisier,GibbonsLavoisier}.}
In contrast to the conformal group, where scale-invariance does imply either a continuous mass spectrum or else zero mass, non-relativistic masses are more like charges
and can assume a discrete set of values (see section~\ref{ageing}).

While this discussion was centred on the free Schr\"odinger equations, see also \cite{Bihlo2017},
the symmetries of non-linear Schr\"odinger equations have been analysed in detail as well, see \cite{Boyer76,Nikitin2001}.

We finish this section with a comment. On p.~\pageref{Niedfoot} we recalled Niederer's result that a Newtonian potential of the form
\begin{equation}
V(t,r) = \frac{g}{\sqrt{t}}\,\frac{1}{r}
\label{funnypot}
\end{equation}
has a dilatation symmetry \cite{Niederer1974}. This generalises as follows. Consider the scaling \cite{ZCEH}
\begin{equation}
t \mapsto t' = \frac{1}{\delta^{2}} t \;\; , \;\;  r \mapsto r' = \frac{1}{\mu}  r\,.
\label{lmscaling}
\end{equation}
and a Newtonian potential $V(t,r)\sim t^{-1/2}r^k$.
We find that the kinetic and potential terms of the usual non-relativistic Lagrangian $L$ scale by the same factor when
\begin{equation}
\mu = \delta^{\frac{3}{2-k}}\,,
\label{mulambda}
\end{equation}
and then
\begin{equation}
L \mapsto \delta^{-\frac{4k-2}{2-k}}L
\quad\Rightarrow\quad
\int L \:\D t  \mapsto \delta^{-\frac{2k+2}{2-k}}\int L \: \D t\,.
\label{actionscaling}
\end{equation}
In the time-distorted Kepler case $k=-1$ (only), namely for the potential
\eqref{funnypot}, we have
\begin{equation}
\mu  = \delta
\quad\Rightarrow\quad
L \to \delta^{2} L \quad\Rightarrow\quad \int L \:\D t \to \int L \: \D t\,.
\label{tkeplerscaling}
\end{equation}
which reproduces Niederer's result for the specific time-dependent potential \eqref{funnypot}.
More generally, we see that the dilatation \eqref{lmscaling}, together with \eqref{mulambda},
is a \emph{symmetry} for the more general potential $V(t,r)\sim t^{-1/2}r^k$ and generalises \cite{Niederer1974}.

The consistency with the approach of section \ref{sec:Eisenhart} is seen for $k=-1$ as follows: extend first the Schr\"odinger
dilation into
\begin{equation}
(t,s,\bx) \mapsto (\delta^{-2}t,s,\delta^{-1}\bx)\,.
\label{sdilat}
\end{equation}
Then all three terms of the Bargmann metric  \eqref{brg} 
\beq
2\D t \D s - 2\frac{1}{\sqrt{t}}\,\frac{1}{r} \D s^2 + \D x^i \D x^j
\label{funnymmetric} \nonumber
\eeq
are multiplied by $\delta^{-2}$. On the other hand, motions in non-relativistic space-time
are the projections of \emph{null} geodesics in the Bargmann space, and the latter are invariant under the rescaling.
Hence one has indeed a symmetry.\footnote{These results are in contrast with \cite{DGH91}
which state that the Bargmann metric for the potential proposed by Dirac \cite{Dirac1/t},
$V_{\rm Dirac}(t,r)= \frac{1}{t}\,\frac{1}{r}$
is also conformally related to the usual Newtonian potential $r^{-1}$. A similar scaling argument explains Kepler's Third Law \cite{DuvalThesis}.
}

\section{Local scale-invariance} \label{sec:lsi}

As a consequence of much interest into equilibrium phase-transitions since the 1950s, conformal invariance was identified  \cite{Polyakov1970}
as a key ingredient for the calculation of co-variant $n$-point correlation functions,\footnote{Relabelling one of the spatial directions as `time',
a conformally covariant two-point function is
$\bigl\langle \phi_1(t_1,\vec{r}_1)\phi_2(t_2,\vec{r}_2)\bigr\rangle = \delta_{x_1,x_2} \bigl( (t_1-t_2)^2 + (\vec{r}_1-\vec{r}_2)^2 \bigr)^{-x_1}$,
where $x_{1,2}$ are the scaling dimensions of the two scaling operators $\phi_{1,2}$. The only consequence of conformal invariance, beyond scale-invariance,
is the constraint $x_1=x_2$. `Covariance' means quasi-primary correlators \cite{BPZ84}.}
especially in two spatial dimensions where the conformal algebra becomes infinite-dimensional and much stronger results hold than for $d>2$ \cite{BPZ84}.
In this section, we give a brief and compact review on whether Schr\"odinger transformations (or suitable extensions)
can be considered similarly as generic space-time transformations such that $n$-point functions are fixed from the requirement of co-variance.
To describe this, a more systematic notation is useful, see table~\ref{tab:gen}. The Schr\"odinger algebra
$\mathfrak{sch}(d)=\bigl\langle X_{\pm 1,0}, {Y}_{\pm 1/2}^{(j)}, M_0, R^{(jk)}\bigr\rangle$ is written compactly as
\BEA
\bigl[ X_n, X_{n'} \bigr] &=& (n-n') X_{n+n'} \nonumber \\
\bigl[ X_n, {Y}_{m}^{(j)} \bigr] &=& \left( \frac{n}{2} - m \right) {Y}_{n+m}^{(j)} \nonumber \\
\bigl[ Y_{m}^{(j)}, Y_{m'}^{(k)} \bigr] &=& \delta_{j,k} \left( m - m'\right) M_{m+m'}  \label{7.1} \\
\bigl[ Y_m^{(j)}, R^{(k\ell)} \bigr] &=& \delta_{j,\ell} Y_{m+n}^{(k)} - \delta_{j,k} Y_{m+n}^{(\ell)}  \nonumber \\
\bigl[ R^{(jk)},R^{(\ell i)}\bigr] &=& \delta_{k,\ell}R^{(ji)} + \delta_{j,i}R^{(k\ell)} - \delta_{k,i}R^{(j\ell)} - \delta_{j,\ell}R^{(ki)} \nonumber
\EEA
with $i,j,k,\ell=1,\ldots,d$. The generators in table~\ref{tab:gen} are for scalars under spatial rotations, see \cite{Hagen:1972pd} for generalisations to higher spin.
As a first consequence, eq.~(\ref{7.1}) immediately suggests an extension to the infinite-dimensional
{\em Schr\"odinger-Virasoro algebra} \cite{Hen94} where $n,n'\in\mathbb{Z}$, $m,m'\in\mathbb{Z}+\frac{1}{2}$.
An explicit representation of (\ref{7.1}) through space-time
transformation is (with $\vec{r}=(r_1,\ldots,r_d)$ and $\partial_{r_j}=\frac{\partial}{\partial r_j}$)
\BEA
X_n &=& - t^{n+1}\partial_t -\frac{n+1}{2} t^n \vec{r}\cdot \vec{\partial}_{\vec{r}} - \frac{\cal M}{4}n(n+1) \vec{r}^2 - \frac{x}{2} t^{n+1}
\nonumber\\
Y_m^{(j)} &=& - t^{m+1/2} \partial_{r_j} - \left( m+\demi\right) {\cal M} t^{m-1/2} r_j
\label{7.2} \\
M_n &=& - t^n {\cal M}
\nonumber \\
R^{(jk)} &=& - \left( r_j \partial_{r_k} - r_k \partial_{r_j}\right) \:=\: - R^{(kj)}
\nonumber
\EEA
where $x$ is the scaling dimension and $\cal M$ the mass of the field (assumed scalar under spatial rotations) these generators act on.
A central charge, of the familiar Virasoro form, can only occur in the commutator $\bigl[ X_n, X_{n'} \bigr]$ \cite{Hen94,UnterbergerRoger2011}.

\begin{table}[tb]
\begin{tabular}{|lclcl|} \hline
$H$   & = & $-\partial_t$                                                                              & = & $X_{-1}$  \\
$D$   & = & $-t\partial_t - \frac{1}{2} \vec{r}\cdot \partial_{\vec{r}} - \frac{x}{2}$                 & = & $X_0$ \\
$K$   & = & $- t^2 \partial_t - t \vec{r}\cdot \partial_{\vec{r}}  -x t  - \frac{\cal M}{2} \vec{r}^2$ & = & $X_1$ \\
$P_j$ & = & $- \partial_{r_j}$                                                                         & = & $Y_{-1/2}^{(j)}$ \\
$B_j$ & = & $- t \partial_{r_j} - {\cal M} r_j$                                                        & = & $Y_{1/2}^{(j)}$  \\
$M$   & = & $-{\cal M}$                                                                                & = & $M_0$   \\
$R^{(jk)}$ & = & $-r_j\partial_{r_k} + r_k \partial_{r_j}$                                             & = & $-R^{(kj)}$ \\ \hline
\end{tabular}
\caption[tabgen]{Different notations for the generators of the Schr\"odinger algebra in $d$ spatial dimensions ($j,k=1,\ldots,d$)
and their definitions as space-time transformations.
\label{tab:gen}}
\end{table}

Together with the conformal algebra, (\ref{7.1}) forms the basis for the construction of more general space-time transformations with a prescribed
dynamical scaling behaviour.\footnote{From dynamical scaling with $z=2$ and time-/space-translation-invariance alone
$\bigl\langle \phi_1(t_1,\vec{r}_1)\phi_2(t_2,\vec{r}_2)\bigr\rangle = (t_1-t_2)^{-(x_1+x_2)/2} \Phi\bigl(\frac{(\vec{r}_1-\vec{r}_2)^2}{t_1-t_2}\bigr)$ with
a non-trivial scaling function $\Phi$.}
A simple way to specify this uses the following axioms \cite{Henkel02} (for simplicity let $d=1$, but generalisations to $d>1$ appear straightforward)
\begin{enumerate}
\item The `time' coordinate transforms as $t\mapsto \frac{\alpha t +\beta}{\gamma t +\delta}$ with $\alpha\delta-\beta\gamma=1$.
The corresponding generators obey $\bigl[ X_n, X_{n'}\bigr]=(n-n')X_{n+n'}$.
\item Dilatations have the generator $X_0 = -t\partial_t - \frac{1}{\theta}r\partial_r - \frac{x}{\theta}$,
where $\theta$ is the {\em anisotropy exponent}.\footnote{For conformal invariance $\theta=1$, for Schr\"odinger-invariance $\theta=2$.}
Time-translations are generated by $X_{-1}=-\partial_t$.
\item Admit spatial translations, with the generator $-\partial_r$.
\item Add further terms, to express the transformation of scaling operators on which these generators are supposed to act.\footnote{Galilei-invariant theories
are important examples on how to specify such terms \cite{Leblond,Sudarshan74}.}
\end{enumerate}
These assumptions are sufficient to obtain a finite list of possibilities which can be stated as follows.
The first result gives the form of $X_n$ as space-time transformations.

\noindent
{\bf Theorem 1:} \cite{Henkel02}  {\it Consider the generators, of first order in $\partial_t$ and $\partial_r$}
\BEQ \label{3:XIIgen}
X_n = - t^{n+1} \partial_t - a_n(t,r) \partial_r - b_n(t,r)
\EEQ
{\it where}
\BEA
a_n(t,r)
&=& \left( \frac{n+1}{\theta}t^n r +
\frac{1}{2}n(n+1)A_{1} t^{n-1} r^{\theta+1} \right)
\left( 1 - \frac{A_{2}}{\theta A_{1}^2}\right)
\nonumber \\
& & + \frac{A_{2}}{(\theta A_{1})^3} t^{n+1}r^{1-\theta}
\left[ \left(1+\theta A_{1}r^{\theta}/t\right)^{n+1} -1\right]
\EEA
{\it and}
\BEA
b_{n}(t,r) &=& \frac{n+1}{\theta} x t^n
+\frac{n(n+1)}{2} t^{n-1}r^{\theta} B_{1}
\left( 1 - \frac{A_{2}}{\theta A_{1}^2}\right)
+ n t^n \frac{A_{2}B_{1}}{\theta^2 A_{1}^3}
\left(1+\theta A_{1} r^{\theta}/t\right)^n
\nonumber \\
& &+ t^n \frac{A_{1}B_{2} -2A_{2}B_{1}}{\theta^2 A_{1}^3}
\left[ (n+1)+(n-1)\left(1+\theta A_{1} r^{\theta}/t\right)^n\right]
\nonumber \\
& &+ t^{n+1} r^{-\theta} \frac{2A_{1}B_{2}-3A_{2}B_{1}}{\theta^3 A_{1}^4}
\left[ 1-\left(1+\theta A_{1} r^{\theta}/t\right)^n\right]
\EEA
{\it and such that one of the following conditions}
\begin{subequations} \label{3:CondiAB}
\begin{align}
(1.)~  &  A_{1} \ne 0 \;\; , \;\; A_{2} = \theta A_{1}^2 \;\; , \;\; B_{1}\ne 0 \;\; , \;\; B_{2}\ne 0
\\
(2.)~  &  A_{1} = A_{2} = 0 \;\; , \;\;  B_{1}\ne 0 \;\; , \;\; B_{2}\ne 0
\\
(3.)~  &  A_{1} \ne 0 \;\; , \;\; A_{2} = 0 \;\; , \;\;  B_{1} \ne 0 \;\; , \;\; B_{2} = 0
\\
(4.)~  &  A_{1}= 0 \;\; , \;\; A_{2} \ne 0 \;\; , \;\; B_{1} = 0 \;\; , \;\; B_{2} \ne  0
\end{align}
\end{subequations}
{\it holds. They are the most general form admitted by axioms 1 \& 2 which satisfy the commutation relations
$[X_{n},X_{n'}]=(n-n')X_{n+n'}$ for all $n,n'\in\mathbb{Z}$.}

Next, one finds the generators related to spatial translations, in order to include axioms~3 and~4. Let first $\theta=2/N$,
then set $m=-\frac{N}{2}+k$ with $k\in\mathbb{Z}$ and let
\BD
Y_m = Y_{k-N/2} = -\frac{2}{N(k+1)}\left( \frac{\partial a_k}{\partial r}\partial_r + \frac{\partial b_k}{\partial r}\right)
\ED
For spatial translations $Y_{-N/2}=-\partial_r$.
The constant $B_1$ from theorem~1 is considered arbitrary. \\

Statements on complete algebras of dynamical symmetries can be given as follows, but only for the anisotropy exponent $\theta=1$ or $\theta=2$ can be Ward identities be specified directly from the symmetry generators.

\noindent {\bf Theorem 2:} \cite{Henkel02}
{\it With the functions $a_n$ and $b_n$ as in theorem~1, $n,n'\in\mathbb{Z}$ and $m=-N/2+k$ with $k\in\mathbb{Z}$, the commutators
\BEQ \label{3:XYComm_II}
\left[ X_n , X_{n'} \right] = (n-n') X_{n+n'} \;\; , \;\;
\left[ X_n , Y_m \right] = \left( n \frac{N}{2} - m \right) Y_{n+m}
\EEQ
hold in one of the following three cases:\\
(i) $A_{1}=A_{2}=B_{2}=0$ and $N$ arbitrary (but $N\neq 1,2$).  \\
(ii) $A_{10}=A_{20}=0$ and $N=1$. For $B_2=0$, there is a further set of generators $Z_n$ with
$n\in\mathbb{Z}$ and\footnote{For $B_2\ne 0$, there are three families $Z_n^{(i)}$, $i=0,1,2$ of generators which close into a Lie algebra.}
the non-vanishing additional commutators are
\BEQ
\left[ Y_m, Y_{m'} \right] = (m-m') B_{1} Z_{m+m'}\;\; , \;\;
\left[ X_n, Z_{n'}\right] = -n'\, Z_{n+n'}
\EEQ
(iii) $A_{2}=A_{1}^2$,  $B_{2}=\frac{3}{2}A_{1} B_{1}$ and $N=2$.
Then for all $n,n'\in\mathbb{Z}$}
\BEQ
\left[ X_n , X_{n'} \right] = (n-n') X_{n+n'} \;\; , \;\;
\left[ X_n , Y_{n'} \right] = (n-n') Y_{n+n'} \;\; , \;\;
\left[ Y_n , Y_{n'} \right] = A_{1} (n-n') Y_{n+n'}
\EEQ

Several comments are in order. \\
{\bf 1.} If $B_1=0$, then $\bigl[ Y_m, Y_{m'}\bigr]=0$ and (\ref{3:XYComm_II}) is a closed Lie algebra,
first identified by Negro, del Olmo and Rodr\'{\i}guez-Marco \cite{Negro97}
and of which the special case $d=1$ is given in Theorem~2.\footnote{Besides spatial rotations, in \cite{Negro97} there is a further purely spatial scaling generator
$D_s$.}  It is nowadays referred to as {\em $\ell$-conformal Galilei algebra}.\footnote{A lot of work has been dedicated to this class of algebras, focusing on possible
central extensions and invariant equations \cite{Aizawa15,Aizawa16,Aizawa16b,Aizawa19,Masterov23}
or physical realisations in the setting of the Pais-Uhlenbeck oscillator model \cite{Galajinsky:2013jma,Galajinsky2015,Andr14,Krivonos16} or fluid mechanics \cite{Snegirev2023}.}
However, for $B_1=0$ the generators do not contain the terms which through Ward identities would describe the transformation
of scaling operators. Then one rests with space-time transformations which cannot be used to constrain $n$-point functions through their
co-variance.
The construction of such terms requires $B_1\ne 0$ but has not yet been solved satisfactorily, whenever $N\ne 1,2$.
For generic $N$, it has been tried to use fractional (Riemann-Liouville) derivatives.

For $N$ integer, the $X_n$ and $Y_m$ act as dynamical symmetries of the Schr\"odinger operator
$\mathscr{S} = -\alpha\partial_t^N + \frac{N^2}{4} \vec{\partial}_{\vec{r}}\cdot\vec{\partial}_{\vec{r}}$, in the sense of Niederer \cite{Niederer:1972zz,Niederer1974}.
Specifically, for $N=4$, hence $\theta=\frac{1}{2}$, one obtains a candidate for a local scaling symmetry at the so-called Lifshitz point
for spin systems with axial next-nearest neighbour interactions. Then the universal form of the scaling function $\Phi$ of two-point correlators is found.
Exact results for spin-spin correlators in the ANNNS model \cite{HenkelPRL97} and numerical simulations in the $3D$ ANNNI model \cite{Pleimling01}
agree with these predictions.\footnote{A field-theoretic second-order $\vep$-expansion gives
\cite{ShpotDiehl2001} $\theta = \frac{1}{2} - 0.0054 \vep^2 + {\rm O}(\vep^3)$ in the ANNNI model in $4.5-\vep$ dimensions.
If that result should be stable under an eventual re-summation of the asymptotic $\vep$-series, at the very least $N\simeq 4$ holds only approximately,
although the changes are probably smaller than numerical error bars in existing simulations.\\
Later work \cite{Rutkevich2011} tried to refute local scale-invariance as sketched above. Therein, however, a different fractional derivative than in \cite{Henkel02}
was used and the authors also did not consider the detailed analysis of the form of the scaling function $\Phi$ through the calculation of moments
\cite{Pleimling01,HenkelPleimling2010}. Instead, they postulate another version of local scale-invariance, of their own making, and promptly refute it.}

\noindent
{\bf 2.} The case (ii) in theorem~2 is the Schr\"odinger-Virasoro algebra. \\
{\bf 3.} Case (iii) gives for the generic situation $A_1\ne 0$ a Lie algebra isomorphism with the conformal algebra, but in a representation
which does not conserve angles and is {\em not} conformal.
While for a long time considered as a mere curiosity, it was only understood recently that this rather represents a new type
of symmetry, the so-called {\em meta-conformal algebra} \cite{Henkel-meta2}, which arises in systems with a directional bias in space.
Their name alludes to the Lie algebra isomorphism with the standard {\em ortho-conformal} Lie algebra in $(1+1)$ space-time dimensions.
In contrast to ortho-conformal invariance, in $(1+2)$ space-time dimensions
there is an infinite-dimensional meta-conformal algebra, isomorphic to the direct sum of {\em three}
Virasoro algebras.\footnote{The same algebra also arises as dynamical symmetry
of $1D$ spatially non-local erosion models \cite{KrugMeakin,Krug,Henkel2016,Henkel2017}.} On the other hand, for $A_1=0$ one is back to the conformal Galilean algebra.
More will be said on this latter algebra in section~\ref{CGA}.

Remarkably, these same infinite-dimensional Lie algebras can be recovered from the Newton-Cartan structures when an arbitrary anisotropy exponent
$\theta$ is admitted \cite{DH-NC}. These two symmetries arises when considering the geodesics of the Newton-Cartan structure. The end result is, in flat space-time
\BEA \label{eq:geodesic}
\mbox{\rm\small time-like geodesic}  &\theta=2 & \mbox{\rm\small Schr\"odinger algebra}  \nonumber \\
\mbox{\rm\small light-like geodesic} &\theta=1 & \mbox{\rm\small conformal Galilean algebra}
\EEA
This analysis centres on the space-time coordinate transformations and does not consider any Ward identities.

\begin{figure}[tb]
\includegraphics[width=0.5\hsize]{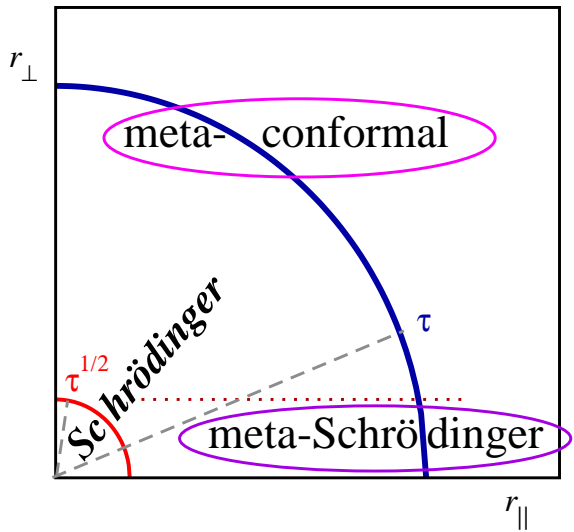}
\caption[fig2]{Spatial regions where various un-biased or biased symmetries can be realised. When distances scale isotropically with time as $r\sim \tau^{1/2}$,
the space-time dynamical symmetry is the Schr\"odinger algebra. If a bias occurs and distances scale in the preferred direction as $r_{\|}\sim \tau$ while
$r_{\perp}\sim \tau^{1/2}$ in the transverse direction, meta-Schr\"odinger invariance is realised. But if $r_{\|}\sim r_{\perp}\sim\tau$, meta-conformal
symmetry may be realised, under certain conditions.
\label{fig2} }
\end{figure}

Table~\ref{tab1} summarises, besides the (ortho-)conformal algebra in $(1+1)D$, the coordinate-transformations for the infinite-dimensional
conformal Galilean and Schr\"o\-din\-ger-Vi\-ra\-so\-ro groups. If applicable, an example of a Schr\"odinger operator on which these transformations act as dynamical
symmetries, is indicated. In addition, it has been understood recently that if a directional bias occurs in the system, the dynamical symmetry can be modified
\cite{Henkel-meta2,Stoimenov22-metaS}.
For example, if a bias is applied to a Schr\"odinger-invariant system along a preferred coordinate $r_{\|}$
and if one uses spatially anisotropic scaling such that for large
distances $r_{\|},r_{\perp}\gg 1$ and large time separations $\tau\gg 1$ and keeps $r_{\|}/\tau$ and $r_{\perp}/\tau^{1/2}$ fixed, the dynamical symmetry turns into the
{\em meta-Schr\"odinger} symmetry. However, if the scaling is made such that $r_{\|}/\tau$ and $r_{\perp}/\tau$ are kept fixed, and certain conditions on sufficiently
long-ranged initial correlators are met, one may rather obtain the {\em meta-conformal} dynamical symmetry. This is illustrated in figure~\ref{fig2}, where the
domains of meta-Schr\"odinger and meta-conformal symmetries are indicated. Both only occur at considerably larger spatial separations than Schr\"odinger symmetry.
This has been checked through exact calculations in the biased Glauber-Ising and spherical models \cite{Henkel-meta2,Henkel23}.

Applications of local scale-invariance in the context of dynamics far from equilibrium and physical ageing will be discussed in section~\ref{ageing}.

\section{Conformal Galilean Algebra}
\label{CGA}

Since Lorentz and Einstein it is well-known that the Galilei group can be obtained from a contraction of the Poincar\'e group,
in the non-relativistic limit when the speed of light $c\to\infty$.
Can one obtain the Schr\"odinger group analogously from a contraction of the conformal group~?
The question was apparently raised first by Barut \cite{Barut} who stated that
\begin{quote}
\textit{\narrower The Schr\"odinger group [arises] from the conformal group by a combined process of contraction and a 'transfer' of the
transformation of mass to the co-ordinates.}
\end{quote}
but he does not define what he means by `transfer'. Since this idea is very interesting,
we shall give a mathematically clean presentation of the argument, but shall also find
that the meaning of `conformal group' suffers a slight modification and that
in the non-relativistic limit one obtains an algebra different from the Schr\"odinger algebra.
We shall follow the presentation given in \cite{HenkelUnterberger}.

\begin{figure}[tb]
\includegraphics[width=0.85\hsize]{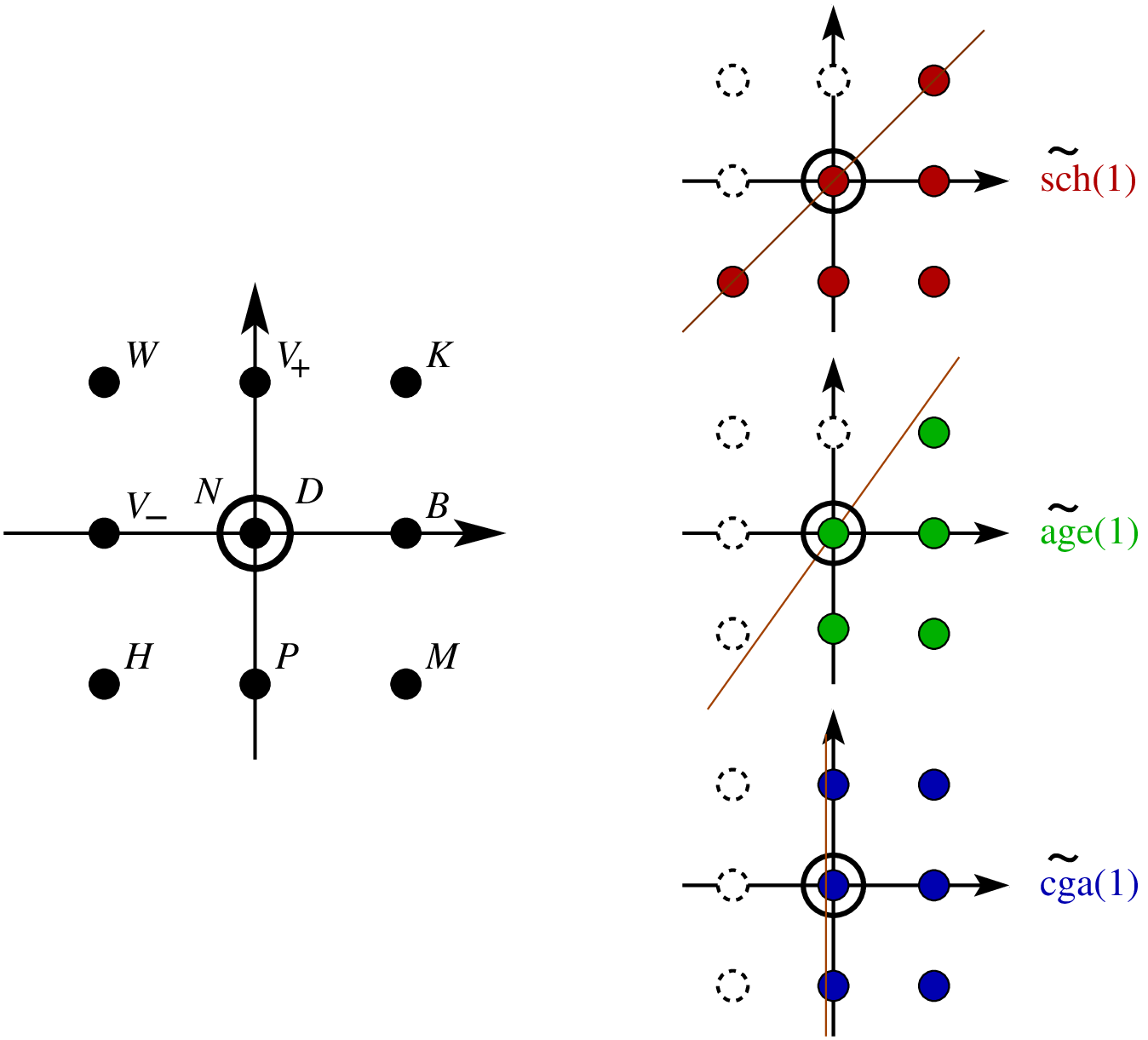}
\caption[fig4]{Left: Root diagram of the complex Lie algebra $B_2$, with the generators $H,P,M,D,B,K$
of table~\ref{tab:gen} and the four additional ones $V_{\pm},W,N$.
Right: the three minimal standard parabolic sub-algebras
$\wit{\mathfrak{sch}}(1)=\mathfrak{sch}(1)\oplus\mathbb{C}N$, $\wit{\mathfrak{age}}(1)=\mathfrak{age}(1)\oplus\mathbb{C}N$
and $\wit{\mbox{\sc cga}}(1)=\mbox{\sc cga}(1)\oplus\mathbb{C}N$.
\label{fig4} }
\end{figure}

As Barut \cite{Barut}, we begin with the massive Klein-Gordon equation
\BEQ \label{8.1}
\left( \frac{1}{c^2} \frac{\partial^2}{\partial t^2}
+ \frac{\partial}{\partial \vec{r}}\cdot \frac{\partial}{\partial \vec{r}}
- {\cal M}^2 c^2 \right) \vph(t,\vec{r}) = 0
\EEQ
Barut now attempted a change of variables via $\partial_t \mapsto {\cal M}c + \frac{1}{c}\partial_t$
but is forced to an ill-defined `transfer'. To implement his idea, we admit
the mass $\cal M$ as a further variable \cite{Giulini:1995te} such that $\vph= \vph_{\cal M}(t,\vec{r})$
and then define a new wave function $\chi$ via
\BEQ
\vph_{\cal M}(t,\vec{r}) = \frac{1}{\sqrt{2\pi\,}\,} \int_{\mathbb{R}} \!\D u\: e^{-\II u {\cal M}} \chi(u,t,\vec{r})
\EEQ
which requires as a necessary condition that $\lim_{u\to\pm\infty} \chi(u,t,\vec{r})=0$. Then eq.~(\ref{8.1}) becomes
\BEQ \label{8.3}
\left( \frac{1}{c^2} \frac{\partial^2}{\partial t^2} + \frac{\partial}{\partial \vec{r}}\cdot \frac{\partial}{\partial \vec{r}}
                                                     + c^2 \frac{\partial^2}{\partial u^2}\right) \chi(u,t,\vec{r}) = 0
\EEQ
which is a massless Klein-Gordon equation in $d+2$ dimensions (and not a massive Klein-Gordon equation in $d+1$ space-time dimensions).
In the new coordinates $\xi_{-1}=\frac{u}{c}, \xi_0=c t$ and $\xi_j=r_j$
with $j=1,\ldots,d$ and $\Psi(\vec{\xi})=\chi(u,t,\vec{r})$, eq.~(\ref{8.3}) becomes $\partial_{\mu}\partial^{\mu} \Psi(\vec{\xi})=0$.
Its dynamical symmetry is the conformal group in its usual form, with generators
\BEA
P_{\mu}    &=& \partial_{\mu} \nonumber \\
M_{\mu\nu} &=& \xi_{\mu}\partial_{\nu} - \xi_{\nu}\partial_{\mu} \nonumber \\
K_{\mu}    &=& 2 \xi_{\mu} \xi^{\nu} \partial_{\nu} -\xi_{\nu}\xi^{\nu}\partial_{\mu} + 2 x\xi_{\mu} \label{8.4} \\
{\cal D}   &=& \xi^{\nu}\partial_{\nu} +x \nonumber
\EEA
with summation convention over repeated indices $\mu,\nu=-1,0,1,\ldots,d$ and the scaling dimension $x$. To prepare the contraction, let
\BEQ
\psi(\zeta,t,\vec{r}) = \chi(u,t,\vec{r}) \mbox{\rm ~~~where~~~} \zeta = u + \II c^2 t
\EEQ
which we believe is the sort of `transfer' Barut might have had in mind.
Finally, to take the non-relativistic limit rewrite (\ref{8.3}) as follows
\BEQ
\left( 2\II \frac{\partial^2}{\partial \zeta \partial t} + \frac{\partial}{\partial \vec{r}}\cdot \frac{\partial}{\partial \vec{r}} \right) \psi(\zeta,t,\vec{r})
= \frac{1}{c^2} \frac{\partial^2}{\partial t^2} \psi(\zeta,t,\vec{r}) = {\rm O}\bigl( c^{-2}\bigr)
\EEQ
which reduces to the free Schr\"odinger equation in the $c\to\infty$ limit.
It remains to write the generators (\ref{8.4}) in this limit, which we do here for $d=1$ for simplicity. We find (and use the notations of table~\ref{tab:gen} and figure~\ref{fig4}).
{\small\BEQ \label{gl:contract}
\begin{array}{lll}
P_{-1} \Psi = -\II c M \psi                             & P_0 \Psi = c M \psi + {\rm O}\bigl( c^{-1}\bigr)            & P_1 \Psi = - P \psi \\
M_{01}\Psi = -c B \psi + {\rm O}\bigl( c^{-1}\bigr)     & M_{-11}\Psi = \II c B \psi + {\rm O}\bigl( c^{-1}\bigr)     & M_{-10}\Psi = \II N \psi + {\rm O}\bigl( c^{-2}\bigr) \\
K_{-1}\Psi = 2\II c K \psi + {\rm O}\bigl( c^{-1}\bigr) & K_0 \Psi = -2c K \psi + {\rm O}\bigl( c^{-1}\bigr)          & K_{1}\Psi = - V_+ \psi + {\rm O}\bigl( c^{-2}\bigr)
\end{array}
\EEQ}
for translations, rotations and expansions, respectively, while for the dilatation
${\cal D}\Psi = (-2X_0+N)\psi$. Herein, we used the further notations
$M=P_{-1}+\II P_0=\frac{1}{c}\partial_t$, $N=-\II M_{-10}=\zeta\partial_{\zeta}-t\partial_t$
and $V_+=-\big(\II/2\bigr)^{1/2}K_1$.
The generators of the Schr\"odinger algebra are given in
table~\ref{tab:gen}.\footnote{Admitting $\cal M$ as a further variable \cite{Giulini:1995te}
and after a Fourier transformation with respect to $\cal M$ in order to introduce the dependence on $\zeta$.}

In order to understand the meaning of these result we give in figure~\ref{fig4}, via a root diagram,
an overview of the commutator relations of the conformal algebra in $(1+1+1)$ dimensions, which is
isomorphic to the complex Lie algebra $B_2$ \cite{Knapp1986}. This illustrates that each generator
$\mathscr{X}\in B_2$ can be linked to a root vector
$\mathfrak{x}$.\footnote{If $\mathfrak{x}_1+\mathfrak{x}_2=\mathfrak{x}_3$, then
$\bigl[\mathscr{X}_1,\mathscr{X}_2\bigr]=\mathscr{X}_3$,
up to a constant factor and up to a linear combination of the roots of the Cartan sub-algebra
$\mathfrak{h}=\bigl\langle D,N\bigr\rangle$. But if $\mathfrak{x}_1+\mathfrak{x}_2$ falls outside
the root diagram, then $\bigl[\mathscr{X}_1,\mathscr{X}_2\bigr]=0$. Each convex set leads to a Lie sub-algebra.
Lie algebras with different root diagrams, up to Weyl transformations, are not isomorphic \cite{Knapp1986}.}
For clarity, we repeat on the right of figure~\ref{fig4} the root diagram
for the Schr\"odinger algebra $\wit{\mathfrak{sch}}(1) = \mathfrak{sch}(1)\oplus \mathbb{C}N$.
Now, the contraction procedure (\ref{gl:contract}) leads to a different algebra, called $
\mbox{\sc cga}(1)$, whose root diagram is also indicated on the right
of figure~\ref{fig4}. Therefore, we have in the non-relativistic limit a projection \cite{HenkelUnterberger}
\BEQ
B_2 \cong \mathfrak{conf}(3) \to \mbox{\sc cga}(1) \not\cong \mathfrak{sch}(1)
\EEQ
In conclusion, Barut's insightful idea \cite{Barut} indeed works, although it leads to a different result than expected.\footnote{The dualisation idea \cite{Giulini:1995te}
has another application: by working out the $n$-point function in dual space, before back-transforming, the $n$-point functions can be shown to obey causality and therefore must be
interpreted as {\it response functions} and not as correlators \cite{HenkelUnterberger,Henkel14a}. This reasoning can be extended to the conformal Galilean algebra,
dualising here with respect to the rapidities $\gamma_j$, which shows that their $n$-point functions are symmetric as required for  correlators \cite{Henkel2017b,Henkel2020}.}

Figure~\ref{fig4} contains more information. Recall from the representation theory of Lie algebras \cite{Knapp1986}
that a {\em minimal standard parabolic sub-algebra} is
spanned by the Cartan sub-algebra $\mathfrak{h}$ and all positive roots of a complex semi-simple Lie algebra.
In figure~\ref{fig4}, positive roots are all roots with lie
to the right of a straight line (brown) going through the origin, where the Cartan sub-algebra $\mathfrak{h}$ lies.
A formal classification of the minimal standard parabolic sub-algebras of
$B_2$ is given in \cite{HenkelUnterberger}. The result is shown in figure~\ref{fig4},
where the (brown) straight line can have three essentially different slopes and leads
to the follows parabolic sub-algebras (up to isomorphisms generated by the transformations of the Weyl group of $B_2$)
\BEQ
\left\{
\begin{array}{ll}
\wit{\mathfrak{sch}}(1) = \mathfrak{sch}(1)\oplus \mathbb{C}N & \mbox{\rm\small Schr\"odinger algebra} \\
\wit{\mathfrak{age}}(1)=\mathfrak{age}(1)\oplus\mathbb{C}N    & \mbox{\rm\small ageing algebra} \\
\wit{\mbox{\sc cga}}(1)=\mbox{\sc cga}(1)\oplus\mathbb{C}N    & \mbox{\rm\small conformal Galilean algebra}
\end{array} \right.
\EEQ
This should be compared with the algebras of space-time transformations constructed in section~\ref{sec:lsi}.
There, it was seen that both the Schr\"odinger algebra $\mathfrak{sch}(d)$ and
the conformal Galilean algebra $\mbox{\sc cga}(d)$ may arise either from a study of possible space-time transformations
respecting scale-invariance or else from the admissible form
of geodesic curves. We now see that these two algebras are also the two main parabolic sub-algebras
of the complex conformal Lie algebra $B_2$.\footnote{Their common sub-algebra
$\mathfrak{age}(d)$ was thought to be related to physical ageing, because of the absence of the time-translation generator
$H$ \cite{HenkelUnterberger}. Section~\ref{ageing} deals with
applications of Schr\"odinger-invariance to physical ageing.} \\

The relationship of the conformal Galilean algebra with either ortho- or meta-conformal algebras may be illustrated in yet a different way.
In $(1+1)$ space-time dimensions (using a more systematic notation of generators\footnote{In $(1+1)$
space-time dimensions the isomorphism of ortho- and meta-conformal algebras can be seen as follows \cite{Henkel02}.
In complex light-cone coordinates $z=t+\II\mu r$, let $\ell_n = - z^{n+1}\partial_z+\Delta z^n$
and similarly for $\bar{\ell}_n$, where $\Delta, \overline{\Delta}$ are the conformal weights.
The ortho-conformal generators are $X_n = \ell_n + \bar{\ell}_n$ and $Y_n=\II\mu\bigl(\ell_n-\bar{\ell}_n\bigr)$.
The meta-conformal generators are $X_n=\ell_n+\bar{\ell}_n$ and $Y_n=\mu\ell_n$.}
in analogy with table~\ref{tab:gen} for the Schr\"odinger algebra) one has for the ortho- and meta-conformal algebras, respectively, the
commutators (with $n,n'\in\mathbb{Z}$)
{\small\begin{subequations} \label{gl:conformal-alg}
\begin{align}
& \left[ X_n, X_{n'} \right] = (n-n') X_{n+n'} \;\; , \;\;
\left[ X_n, Y_{n'} \right] = (n-n') Y_{n+n'}  \;\; , \;\;
\left[ Y_n, Y_{n'} \right] = -\mu\, (n-n') X_{n+n'} \label{gl:ortho-conf} \\
& \left[ X_n, X_{n'} \right] = (n-n') X_{n+n'} \;\; , \;\;
\left[ X_n, Y_{n'} \right] = (n-n') Y_{n+n'}  \;\; , \;\;
\left[ Y_n, Y_{n'} \right] = \mu\, (n-n') Y_{n+n'} \label{gl:meta-conf}
\end{align}
\end{subequations}
}
where $\mu=1/c$ is related to the speed of light \cite{Henkel-meta2}. The Lie algebra contractions now simply arises in the
$\mu\to 0$ limit which give from (\ref{gl:conformal-alg}) the commutators
\begin{align} \label{gl:conf-gal}
& \left[ X_n, X_{n'} \right] = (n-n') X_{n+n'} \;\; , \;\;
\left[ X_n, Y_{n'} \right] = (n-n') Y_{n+n'}  \;\; , \;\;
\left[ Y_n, Y_{n'} \right] = 0
\end{align}
of the conformal Galilean algebra $\mbox{\sc cga}(1)$. The forms (\ref{gl:conformal-alg}) suggests the possibility of an infinite-dimensional extension, which however is
possible for the ortho-conformal algebra (\ref{gl:ortho-conf}) in $(1+1)$ dimensions only and for the meta-conformal algebra (\ref{gl:meta-conf}) in $(1+1)$ and $(1+2)$ dimensions.
On the other hand, the conformal Galilean algebra not only can be written for any space dimension $d$ but can always be extended to an infinite-dimensional algebra with $n,n'\in\mathbb{Z}$.
An explicit space-time representation of the conformal Galilean generators in $(1+d)$ dimensions is (with $\vec{r}=\bigl(r_1,\ldots,r_d\bigr)$)
\BEA X_n &=&
- t^{n+1}\partial_t - (n+1) t^n \vec{r}\cdot\vec{\partial} -
x (n+1)t^n - n(n+1) t^{n-1} \vec{\gamma}\cdot\vec{r}
\nonumber \\
Y_n^{(j)} &=& - t^{n+1} \partial_{j} - (n+1) t^n \gamma_j  \label{gl:altrep} \\
R_0^{(jk)} &=& - \bigl( r_j \partial_{k} -  r_k \partial_{j} \bigr)
- \bigl( \gamma_j \partial_{\gamma_k}-\gamma_k
\partial_{\gamma_j}\bigr); \qquad j \ne k \nonumber
\EEA
with $\partial_j = \frac{\partial}{\partial r_j}$, $x$ is a scaling dimension,
the spatial rotation generators were included and we also wrote the terms coming from the rapidities $\gamma_j$, $j=1,\ldots,d$.
In $(1+1)$ dimensions, the maximal finite-dimensional sub-al\-ge\-bra is
$\left\langle X_{\pm 1,0},Y_{\pm 1,0}\right\rangle = \left\langle V_+, D,P, K, B, M\right\rangle$, see also figure~\ref{fig4}.

\begin{figure}[tb]
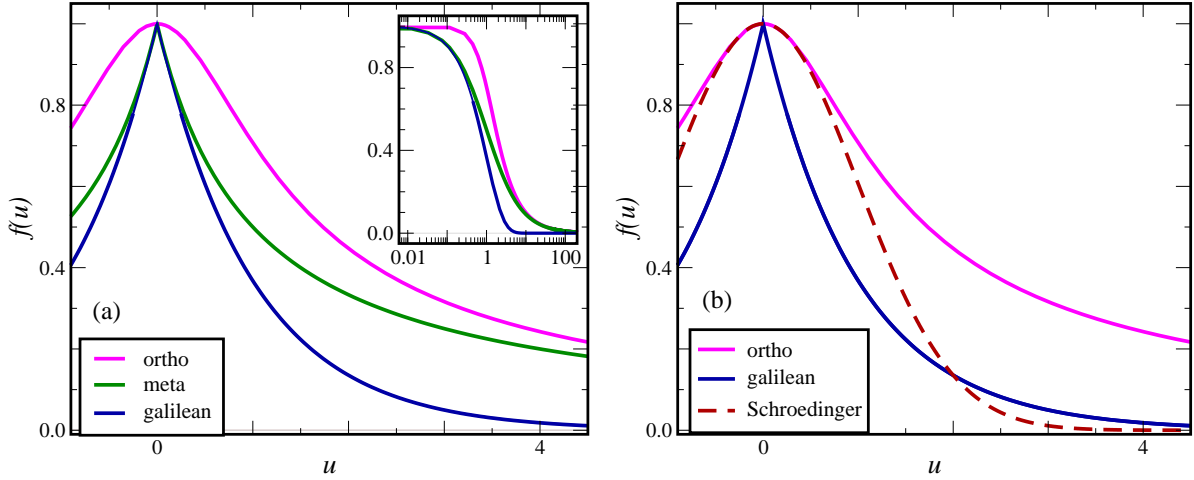

{\includegraphics[scale=0.475]{correlateur1.eps}~~~\includegraphics[scale=0.475]{correlateur2.eps}}
\caption[fig5]{(a) Scaling function $f(u)$ of the covariant two-point correlator ${C}(t,r)=t^{-2x_1}f(r/t)$,
over against the scaling variable $u=r/t$, for the ortho-conformal,
meta-conformal and conformal Galilean algebras, in $(1+1)D$, from eq.~(\ref{gl:correlateurs}).
The inset further underlines the different behaviour for $u\ll 1$ and $u\gg 1$.
(b) Comparison with the scaling function obtained from Schr\"odinger-invariance,
clearly distinct from both ortho-conformal and conformal Galilean invariance.
\label{fig5}}
\end{figure}

The physical difference of these three algebras is further illustrated by the distinct forms of the two-point function
$C(t,r)=\bigl\langle \phi_1(t,r)\phi_2(0,0)\bigr\rangle$, derived from the
condition of co-variance under the maximal finite-dimensional sub-al\-ge\-bra $\bigl\langle X_{\pm 1,0},Y_{\pm 1,0}\bigr\rangle$ \cite{Henkel-meta2}

\BEQ \label{gl:correlateurs}
C(t,r) = \left\{
\begin{array}{ll} (\bigl(t^2 + \mu^2 r^2\bigr)^{-x_1} \exp\bigl( -(2\gamma_1/\mu) \arctan \bigl(\mu |r/t|\bigr)\bigr)  & \mbox{\rm\small ortho-conformal} \\
                  t^{-2x_1} \bigl( 1 + \gamma_1/\mu |r/ t| \bigr)^{-2\gamma_1/\mu}                                     & \mbox{\rm\small meta-conformal} \\
                  t^{-2x_1} \exp\bigl( -2\gamma_1 |r/t|\bigr)                                                          & \mbox{\rm\small conformal Galilean}
\end{array} \right.
\EEQ
where the constraints $x_1=x_2$ and $\gamma_1=\gamma_2$ hold.\footnote{In the $\mu\to 0$ limit,
both ortho- and meta-conformal forms (\ref{gl:correlateurs}) reduce to the
conformal Galilean correlator \cite{Henkel-meta2}. For ortho-conformal invariance,
the conformal weight $\Delta = \bigl(x_1-\II\gamma_1/\mu\bigr)/2$.
Three-point functions $\bigl\langle \phi_1(t_1,\vec{r}_1)\phi_2(t_2,\vec{r}_2)\phi_3(t_3,\vec{r}_3)\bigr\rangle$ can be fixed similarly \cite{HenkelPleimling2010}.}
In contrast with the Schr\"odinger algebras, which predicts co-variant response functions, the
co-variant $n$-point functions found from these three algebras are {\em correlation functions} \cite{Henkel14a}.
The qualitative behaviour of the associated scaling functions is shown in figure~\ref{fig5}a.
For large arguments of the scaling variable $u$, both ortho- and meta-conformal correlators decay algebraically,
whereas the meta-conformal correlator has an exponential decay.
On the other hand, for $u$ small, both meta-conformal and conformal Galilean correlators are not differentiable at $u=0$,
whereas the ortho-conformal correlator has a
rounded form.\footnote{For $(1+2)D$ meta-conformal invariance, the correlator interpolates between the $(1+1)D$
meta-conformal and the ortho-conformal correlator \cite{Henkel-meta2}.}
In figure~\ref{fig5}b, the scaling functions are compared with the one obtained from Schr\"odinger-invariance
(a response function) which clearly highlights their difference ($f(u)\sim e^{-u^2}$ has a Gaussian form for Schr\"odinger-invariance).

\section{Super-selection rules}\label{Lavoisier}
We now revisit the conservation of mass. It is not simply some dynamical symmetry but has deep connections with central extensions of the space-time symmetry algebra.

{\bf 1.} In classical many-particle physics one may use a standard (i.e. non-projective) representation of the Galilei algebra. In an inertial frame, Newton's
equation of motion for a $N$-particle system with positions $\vec{r}_a(t)$ are
\BEQ \label{Lav1}
m_a \ddot{\vec{r}}_a = \vec{F}_a \;\; ~~;~~ \; a = 1,\ldots,N
\EEQ
Herein $m_a$ is the mass of the a$^{\rm th}$ particle and $\vec{F}_a$ is the force acting on it. For an isolated system, $\sum_{a=1}^N \vec{F}_a=\vec{0}$.
Summing over all particles gives
\BD
\frac{\D}{\D t} \left( \sum_{a=1}^N m_a \dot{\vec{r}}_a \right) = \sum_{a=1}^N m_a \ddot{\vec{r}}_a = \sum_{a=1}^N \vec{F}_a = \vec{0}
\ED
which means that the total momentum $\vec{P}$ is conserved
\BEQ \label{Lav2}
\vec{P} = \sum_{a=1}^{N} m_a \dot{\vec{r}}_a = \mbox{\rm cste.}
\EEQ
such that Newton's third axiom has been checked. In textbooks of classical mechanics this is usually derived from spatial translation-invariance.
In addition to the well-known conservation law (\ref{Lav2}), the total mass is also conserved. To see this, change the
inertial frame through a Galilei transformation
\BEQ \label{Lav3}
t \mapsto t' = t \;\; , \;\; \vec{r}_a \mapsto \vec{r}_a' = \vec{r}_a + \vec{v} t
\EEQ
under which (\ref{Lav1}) clearly is co-variant. The momentum conservation (\ref{Lav2}) becomes
\BD
\vec{P}' = \sum_{a=1}^N m_a \dot{\vec{r}}_a' = \sum_{a=1}^N \left( m_a \dot{\vec{r}}_a + m_a \vec{v} \right) = \vec{P} + \sum_{a=1}^N m_a \vec{v}
\ED
Since both momenta $\vec{P}$ and $\vec{P}'$ are constant, one has the further conservation law
\BEQ
\vec{v} \sum_a m_a = \vec{P}' - \vec{P}  = \mbox{\rm cste.} ~~~\Rightarrow~~~ \boxed{ ~\sum_a m_a = M = \mbox{\rm cste.}~ }
\EEQ
since the velocity $\vec{v}$ is arbitrary. Therefore the total mass $M$ of a non-relativistic system is always kept fixed,
which is obtained here from the non-centrally extended representation (\ref{Lav3}).
This mass conservation was established by Lavoisier more than 200 years ago, stating that\footnote{Nothing is created, neither artificially, nor in Nature, and one may pose as a principle that
in all operations there is the same quantity of matter before and after the operation.} \cite{Lavoisier}
\begin{quote}
{\sl ``Rien ne se cr\'ee, ni dans les op\'erations de l'art, ni dans celles de la nature, et l'on peut poser en principe que, dans toute op\'eration,
il y a une \'egale quantit\'e de mati\`ere avant et apr\`es l'op\'eration.''}
\end{quote}
Although it is very important in practise, mass conservation appears here as a circumstantial result, found as a by-product of momentum conservation \cite{GibbonsLavoisier}.

{\bf 2.} This becomes very different when one goes over to non-relativistic quantum mechanics. For a free particle, the entire information is contained in the wave
equation $\psi(t,x)$ which obeys the wave equation (for notational simplicity in $d=1$ space dimensions)
\BEQ \label{Lav5}
\II \hbar \frac{\partial \psi(t,x)}{\partial t} = - \frac{\hbar^2}{2m} \frac{\partial^2 \psi(t,x)}{\partial x^2}
\EEQ
where $m$ is the mass of the particle and $\hbar$ is Planck's constant. While this equation is clearly invariant under temporal and spatial translations, it is
also invariant under the Galilei transformation
\BEQ
t \mapsto t' = t \;\; , \;\; x \mapsto x' = x + v t
\EEQ
but the wave function transforms non-trivially
\BEQ \label{Lav7}
\psi(t,x) \mapsto \psi'(t,x) = \exp\left[\frac{\II}{\hbar} \left( m x v + \frac{1}{2} m t v^2 \right) \right] \psi(t,x-vt)
\EEQ
The importance of such {\em projective representations} was pointed out by Bargmann \cite{Barg54}. For our purposes, it is sufficient to recall that both the wave equation
(\ref{Lav5}) as well as the law of probability conservation
\BEQ
\frac{\partial \rho(t,x)}{\partial t} + \frac{\partial j(t,x)}{\partial x} = 0
\EEQ
transform co-variantly under the projective representation (\ref{Lav7}), whenever $m\ne 0$.
Herein the probability density $\rho$ and the probability current $j$ are given by
\BD
\rho(t,x) = \psi^*(t,x) \psi(t,x) \;\; , \;\;
j(t,x) = \frac{\hbar}{2m\II} \left( \psi^*(t,x) \frac{\partial \psi(t,x)}{\partial x} - \psi(t,x) \frac{\partial \psi^*(t,x)}{\partial x} \right)
\ED
This projective effect in (\ref{Lav7}) cannot be eliminated through a change of variables. It also follows that
for $m\ne 0$, the wave function must be complex-valued.
For a better algebraic understanding, we consider the Lie algebra generator $B$, obtained for infinitesimal $v$ from, (\ref{Lav7})
\BEQ
B = - t \partial_x - \frac{\II m}{\hbar} x \;\; , \;\; P = - \partial_x
\EEQ
along with the generator $P$ of spatial translations. These are already given by Niederer \cite{Niederer:1972zz}.
In contrast to standard representations, their commutator
\BEQ
\bigl[ B, P \bigr] = - \frac{\II m}{\hbar} =: \frac{\II}{\hbar} M
\EEQ
does not vanish for $m\ne 0$. Since $M$ does commute with all other generators of the Galilei algebra,
it provides a {\em central extension} of the (non semi-simple) Galilei algebra.\footnote{For
finite-dimensional Lie algebras $\mathfrak{g}$, central extensions only exist if $\mathfrak{g}$
is not semi-simple. Then central extensions cannot be absorbed into a change of coordinates \cite{UnterbergerRoger2011}.}
The presence of a non-vanishing mass $m\ne 0$ modifies profoundly the underlying mathematical structure.\footnote{See \cite{SudarshanMukunda} for
a classifications of representations of the Galilei group with either $m=0$ or $m\ne 0$, in the context of classical mechanics.}

{\bf 3.} Mass conservation can be seen as a consequence of the central extension and takes a particularly interesting form in many-body systems.
When applying spatial translation-invariance and Galilei-invariance, in the form of the co-variance conditions $P C^{[n]}= B C^{[n]}=0$
with a $n$-point function $C^{[n]}= C(t_1,\ldots,t_n;x_1,\ldots,x_n)$, we find first the reduction
\BD
C^{[n]} = C\bigl( t_1,\ldots,t_n;x_1-x_n,x_2-x_n,\ldots,x_{n-1}-x_n\bigr)
\ED
and furthermore
\BEA
B C^{[n]} &=& \left[ - (t_1-t_n)\partial_{x_1} - \ldots - (t_{n-1}-t_n)\partial_{x_{n-1}} \right. \nonumber \\
& & \left. - \frac{\II}{\hbar} \Bigl( m_1(x_1-x_n) + \ldots + m_{n-1}(x_{n-1}-x_n) \Bigr) \right.
\nonumber \\
& & \left.- \frac{\II}{\hbar} x_n \bigl( m_1 + m_2 +\ldots + m_n\bigr) \right]C^{[n]} = 0
\nonumber
\EEA
Again because of spatial translation-invariance, the correlator $C^{[n]}$ only depends on the differences
$x_1-x_n$, \ldots, $x_{n-1}-x_n$ but cannot depend on $x_n$ alone. Hence one must have
\BEQ \label{Lav11}
\boxed{~\bigl( m_1 + m_2 +\ldots + m_n\bigr) C^{[n]} = 0~}
\EEQ
This a modern rephrasing of Bargmann's result \cite{Barg54}:
{\em a theory which is spatially translation-invariant and Galilei-invariant decomposes into sectors, each with a fixed mass, such
that any $n$-point functions between these sectors vanish.} Since it is a stronger constraint than usual selection rules from internal symmetries,
it is usually called the {\em Bargmann super-selection rule}. Because of (\ref{Lav7}), the complex conjugate $\psi^*$ has a negative mass $m^*=-m<0$ such that the
condition (\ref{Lav11}) can indeed be satisfied. From the present point of view, mass conservation is a fundamental property of a Galilean-invariant theory, rather
than a circumstantial by-product.

{\bf 4.} When studying relaxational phenomena, the field-theoretic descriptions only involve real-valued fields. Certainly, this does not mean that such theories
cannot be Galilei-invariant, but the notion of `complex conjugate' has to be adapted. Indeed, in non-equi\-li\-bri\-um field theory \cite{Jans92,Domi76},
besides the real-valued order-parameter field
$\phi$ one considers another real-valued field, the response field $\wit{\phi}$. In such theories, averages are calculated from functional integrals
$\langle A\rangle = \int \!{\cal D}\phi{\cal D}\wit{\phi}\: A(\phi,\wit{\phi}) e^{-{\cal J}[\phi,\wit{\phi}]}$. For a free particle at temperature $T$, the action reads
\BEQ
{\cal J}[\phi,\wit{\phi}]  = \int \!\D t\D \vec{r}\: \left[ \wit{\phi} \bigl( \partial_t - \Delta_{\vec{r}} \bigr) \phi - T \wit{\phi}^2 \right]
\EEQ
Here the response field acts as `complex conjugate'. If the order parameter $\phi$ has a mass $M>0$, the conjugate response field must have a mass $\wit{M}=-M<0$.
For $N$-particle observables, each field $\phi_j$ of mass $M_j$,
the generators of spatial translations and Galilei transformations can be written as (with $m=\pm\frac{1}{2}$, see table~\ref{tab:gen})
\BEQ \label{B:19}
Y_{m} = \sum_{j=1}^N \left[ - t_j^{m+1/2} \frac{\partial}{\partial x_j} - \left( m+\frac{1}{2} \right) M_j t_j^{m-1/2} x_j \right]
\EEQ
such that their commutator is
\BEQ
\bigl[ Y_{1/2}, Y_{-1/2} \bigr] = -\bigl(M_1+\ldots+M_n\bigr) =: M
\EEQ
We recognise the central extension by the generator $M$ and also read off the Bargmann super-selection rule $M_1 + \ldots + M_n=0$. This means that averages
such as $\langle \phi \wit{\phi}\rangle$, $\langle \phi^2 \wit{\phi}^2\rangle$ and so on can be fixed from their co-variance. This will be explained further in
section~\ref{ageing}.

{\bf 5.} For comparison, we briefly reconsider the same question for the conformal Galilean algebra. {\it A contrario} to the standard Galilean algebra,
accelerations are also present \cite{Lukierski07}. For an $n$-particle system, the Galilean generators are now (with $m=\pm 1,0$)
\BEQ \label{B:21}
Y_m = \sum_{j=1}^n \left(- t_j^{m+1} \frac{\partial}{\partial x_j} - (m+1) \gamma_j t_j^m \right)
\EEQ
We compare the root diagrams in figure~\ref{fig4}. The standard Galilean algebra is spanned by
$\langle B,P,M\rangle \stackrel{\wedge}{=} \langle Y_{\frac{1}{2}},Y_{-\frac{1}{2}}, M\rangle$, see eq.~(\ref{B:19}), where the central extension $M$ was already included.
But the space-transformations of the conformal Galilean algebra are spanned by $\langle K,B,M\rangle \stackrel{\wedge}{=}\langle Y_1, Y_0, Y_{-1}\rangle$
as given by (\ref{B:21}).
Therefore it is clear from figure~\ref{fig4} (or eq.~(\ref{B:21}))
that $\bigl[ Y_{m},Y_{m'}\bigr]=0$ and no central extension exists in this case. For the two-point function $C^{[2]}$, it can be
easily shown, from the co-variance conditions $Y_m C^{[2]}=0$,
that the two `rapidities' are equal: $\gamma_1=\gamma_2$ \cite{HenkelPleimling2010,Bagchi10,Bagchi:2018}. Hence the physical r\^ole of the `masses' $M_j$ and the `rapidities' $\gamma_j$
is different.\footnote{For the infinite-dimensional extension of the conformal Galilean algebra, the usual central extensions of Virasoro form are of course admissible.}
See figure~\ref{fig5} for the comparison of the forms of the two-point scaling functions, according to ortho-conformal, meta-conformal, conformal Galilean and Schr\"odinger invariance.

\section{Physical ageing \label{ageing}}

Galilei-invariance and the Bargmann super-selection rules find a direct application in the context of physical ageing far from equilibrium.
Physical ageing is a typical behaviour of glasses \cite{Stru78,Arce22}.
Here we shall be exclusively interested in the dynamical symmetry principles which are best explained in the ageing of more simple magnetic systems,
without disorder \cite{Cugl03,GodrecheLuck02,HenkelPleimling2010}.

\begin{figure}[tb]
\hspace{-0.8truecm}
\includegraphics[width=.65\hsize]{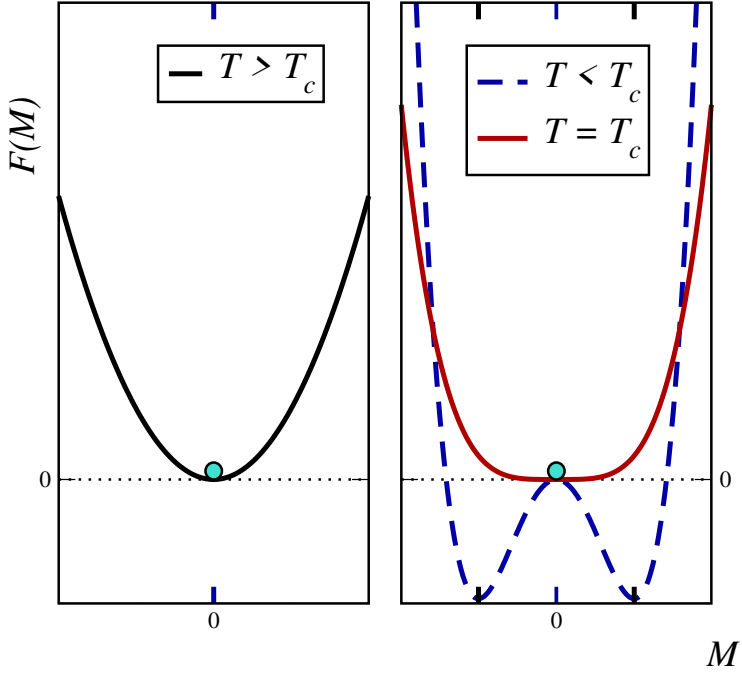}
\caption[fig3]{Schematic free energy before a quench (left panel) and after a quench to either $T=T_c$ or $T<T_c$ (right panel).
The state of the system is symbolised by the small ball.
\label{fig3} }
\end{figure}

Consider a many-body system whose equilibrium state is either critical (with dynamically created long-range correlations) or else has more than one distinct but equivalent
equilibrium states. Roughly speaking, physical ageing arises when the time-evolution starts from an initial state which is different from the
equilibrium state. For example, one might obtain this situation via quenching a system from a fully disordered initial state to a state either onto or else below a
critical temperature $T_c>0$, see figure~\ref{fig3}. After the quench, the system is far from equilibrium, since it is no longer at a stable minimum of the free energy.
Ageing can be monitored through the correlations of the space-time-dependent order-parameter $\phi(t,\vec{r})$. One
measures for instance the single-time correlator or the two-time auto-correlator
\beq
C(t,\vec{r}) = \left\langle \phi(t,\vec{r})\phi(t,\vec{0})\right\rangle \;\; , \;\; C(t,s) = \left\langle \phi(t,\vec{0})\phi(s,\vec{0})\right\rangle
\eeq
where the averages $\bigl\langle \cdot \bigr\rangle$ are over sample histories (and possibly over an ensemble of initial conditions as well)
and for simplicity spatial translation-invariance was assumed. The initial average order-parameter
$\left\langle \phi(0,\vec{r})\right\rangle=0$ is taken to vanish.
By definition, {\em physical ageing} occurs if the following three defining conditions are satisfied \cite{HenkelPleimling2010}
\begin{enumerate}
\item slow relaxational dynamics, not described by a simple exponential with a finite relaxation time
\item breaking of time-translation-invariance
\item dynamical scaling
\end{enumerate}

\begin{figure}[tb]
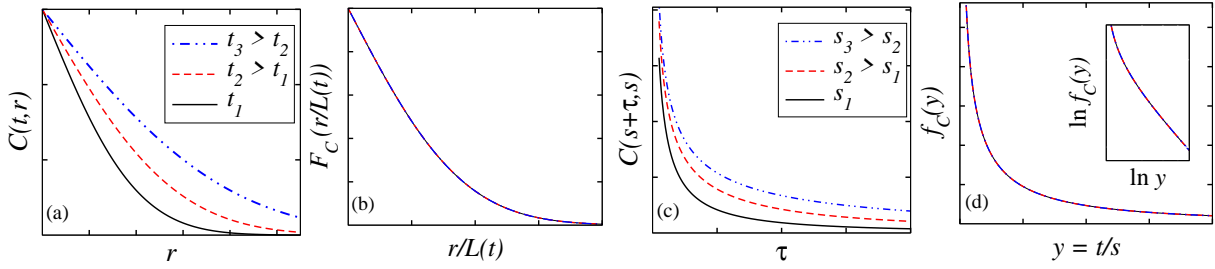

\hspace{-0.8truecm}
\includegraphics[width=.26\hsize]{GI-C1-tdiff-grand.eps}~\includegraphics[width=.26\hsize]{GI-C1-skal-grand.eps}~
\includegraphics[width=.26\hsize]{EW-C2-sdiff-grand.eps}~\includegraphics[width=.26\hsize]{EW-C2-skal-grand.eps}
\caption[fig1]{Illustration of the characteristic data collapse of physical ageing.
Panel (a) shows a typical behaviour of a single-time correlator for different times
$t_3>t_2>t_1$, while (b) shows the collapse onto a single curve when distances $r=|\vec{r}|$
are measured in units of the dynamical length scale $L(t)$.
Panel (c) similarly illustrates the two-time autocorrelator in dependence of $\tau=t-s$,
for different waiting times $s_1<s_2<s_3$
and panel (d) shows that these data collapse when replotted as a function of $y=t/s$.
The log-log plot in the inset shows the asymptotic power-law form $f_C(y)\sim y^{-\lambda/z}$.
\label{fig1} }
\end{figure}

Figure~\ref{fig1} schematically illustrates how ageing can be detected from correlation functions.
The curves of $C(t,\vec{r})$ do depend on time, hence there is
{\em no} time-translation-invariance. But if the same data are replotted over against $\bigl|\vec{r}\bigr|/L(t)$,
where $L(t)\sim t^{1/z}$ is the time-dependent length of the ordered clusters,
($z$ is the dynamical critical exponent) a data collapse occurs.
Similarly, the curves of the two-time autocorrelator $C(t,s)$, when plotted over against the time difference $\tau=t-s$,
do depend on the waiting time $s$ and time-translation-invariance is broken.
Again, when replotted over against $y=t/s$, a data collapse occurs. Hence in the limit of large times, one finds the scaling forms
\beq
C(t,\vec{r}) = t^{-c} F_C\left( \frac{\vec{r}}{L(t)}\right) \;\; , \;\; C(t,s) = s^{-b} f_C\left(\frac{t}{s}\right)
\eeq
The inset in figure~\ref{fig1}d illustrates the generic power-law behaviour of $f_C(y)\sim y^{-\lambda/z}$ for
$y\gg 1$ large. The auto-correlation exponent $\lambda$
is universal but for a non-conserved order-parameter it is independent of all equilibrium critical exponents \cite{Taeu14}.
The renormalisation group asserts that scaling functions such as $F_C$ and $f_C$ are universal.
Then their functional from should only depend on global system properties such as dimension and  global symmetries
but should be independent of most microscopic `details' of a specific Hamiltonian.
Finding their form, independently of studies in specific models, then calls for a convenient dynamical symmetry.

Probably the most simple system with a dynamical exponent $z=2$ is the {\em Edwards-Wilkinson model}, see \cite{BarabasiStanley1995},
for the height $h=h(t,\vec{r})$ of a growing interface.\footnote{Microscopically, this model can be obtained by depositing particles on a surface.
If a particle arrives, it sticks to its point of arrival, but only after having relaxed to the lowest height in the immediate neighbourhood of the arrival point.
The long-range properties of the interface, and their fluctuations, are then described by (\ref{gl:ew}). }
In a frame where the average height is constant, $\partial_t \langle h(t,\vec{r})\rangle =0$, the height fluctuations are described by the Langevin equation
\beq \label{gl:ew}
\partial_t h(t,\vec{r}) = \frac{1}{2\cal M}\Delta_{\vec{r}} h(t,\vec{r}) +  \eta(t,\vec{r})
\eeq
where $\Delta_{\vec{r}}$ is the spatial Laplacian and fluctuations enter through the centred Gaussian white noise with variance
$\left\langle \eta(t,\vec{r})\eta(t',\vec{r}')\right\rangle =  2T {\cal M} \delta(t-t')\delta(\vec{r}-\vec{r}')$.
Following \cite{Henkel-tutoriel}, we shall use this simple model, with its linear Langevin equation,
to illustrate some of the main aspects of dynamical Schr\"odinger symmetry.
Clearly, the noise term in (\ref{gl:ew}) breaks any space-time
symmetry beyond simple translation- and rotation-invariance. Hence the noisy eq.~(\ref{gl:ew}), as it stands, cannot be Schr\"odinger-invariant.

Eq.~(\ref{gl:ew}) can be obtained as a classical equation of motion of the non-equi\-li\-bri\-um Janssen-de Dominicis field theory \cite{Jans92,Domi76,Taeu14},
with the action ${\cal J}[h,\wit{h}] = {\cal J}_0[h,\wit{h}] + {\cal J}_b[\wit{h}]$ decomposed into a deterministic and a noise part, respectively
\BEA
{\cal J}_0[h,\wit{h}] &=& \int \! \D t\D\vec{r}\: \wit{h} \bigl( \partial_t - (2{\cal M})^{-1}\Delta_{\vec{r}} - j \bigr) h \nonumber \\
{\cal J}_b[\wit{h}]   &=& -T \int \! \D t\D\vec{r}\: \wit{h}^2
\label{gl:JD}
\EEA
Herein, $\wit{h}$ is the response field conjugate to the height field $h$.
Averages are computed from the functional integral $\langle \mathscr{A}\rangle = \int {\cal D}h {\cal D}\wit{h}\; \mathscr{A}[h] e^{-{\cal J}[h,\wit{h}]}$.
Notably, one distinguishes
two-particle correlation and response functions \cite{Cugl03,HenkelPleimling2010,Taeu14},
\begin{subequations} \label{gl:def}
\begin{align} \label{gl:Cdef}
C(t_1,t_2;\vec{r}_1,\vec{r}_2) &= \biggl\langle h(t_1,\vec{r}_1)h(t_2,\vec{r}_2)\biggr\rangle \\
R(t_1,t_2;\vec{r}_1,\vec{r}_2) &= \left.\frac{\delta \langle h(t_1,\vec{r}_1)}{\delta j(t_2,\vec{r}_2)}\right|_{j=0}
= \left\langle h(t_1,\vec{r}_1)\wit{h}(t_2,\vec{r}_2)\right\rangle
\label{gl:Rdef}
\end{align}
\end{subequations}
which explains the purpose of the source field $j$ in eq.~(\ref{gl:JD}) and the name of the response field $\wit{h}$.

Now, the deterministic part ${\cal J}_0[h,\wit{h}]$ of the action is Schr\"odinger-invariant (related to the heat equation).
This allows us to identify the following properties of the height field $h$ and its conjugate response field $\wit{h}$:
\begin{center}\begin{tabular}{|lll|}
\noalign{\smallskip}\hline\noalign{\smallskip}
height field  $h$ :~       & scaling dimension $x$       & mass ${\cal M}>0$  \\
response field $\wit{h}$ : & scaling dimension $\wit{x}$ & mass $\wit{\cal M}=-{\cal M}<0$  \\
\noalign{\smallskip}\hline
\end{tabular}\end{center}
It follows from the Bargmann super-selection rules that the $(n+m)$-point deterministic correlator,
computed only with the part ${\cal J}_0[h,\wit{h}]$ of the action, obeys
\beq \label{gl:Barg}
C^{[n,m]}_0 = \left\langle h_1 \ldots h_n \wit{h}_1 \ldots \wit{h}_m\right\rangle_0 = \delta_{n,m} \mathscr{C}^{[n]}
\eeq
such that only deterministic averages with an equal number of $h$- and $\wit{h}$-fields can be non-vanishing.
A non-trivial example would be the response function $R=C^{[1,1]}$,
see (\ref{gl:Rdef}). However, the deterministic correlator $C=C^{[2,0]}_0=0$
vanishes.\footnote{It is a basic textbook result of non-equi\-li\-bri\-um field theory that $\langle \wit{h}_1 \ldots \wit{h}_m\rangle =0$ \cite{Taeu14}.}
A formal expansion \cite{Pico04} in the full action in terms of the `temperature' $T$ then shows from (\ref{gl:Rdef})
that the (noisy) response function $R=\langle h \wit{h}\rangle$
\beq \label{gl:Rred}
R(t,s;\vec{r}) 
= \left\langle h(t,\vec{r}) \wit{h}(s,\vec{0}) e^{-{\cal J}_b[\wit{h}]} \right\rangle_0 =
\left\langle h(t,\vec{r}) \wit{h}(s,\vec{0})\right\rangle_0 = R_0(t,s;\vec{r})
\eeq
which is computed in a stochastic model, is identical to the deterministic response $R_0$ found from Schr\"odinger-invariance. Similarly
the correlator $C=\langle h h \rangle$
\beq \label{gl:Cred}
C(t,s;\vec{r}) 
= \left\langle h(t,\vec{r}) {h}(s,\vec{0}) e^{-{\cal J}_b[\wit{h}]} \right\rangle_0 =
T \int \!\D u \D\vec{R}\: \left\langle h(t,\vec{r}) {h}(s,\vec{0}) \wit{h}^2(u,\vec{R}) \right\rangle_0
\eeq
reduces to an integral of a deterministic three-point response function \cite{Pico04}. The exact reduction formul{\ae} (\ref{gl:Rred},\ref{gl:Cred})
are the basis for finding the scaling functions of responses and correlators.

We note an important feature of Schr\"odinger-invariance: the requirement of co-variance fixes directly {\em response functions},
such as  $R=\bigl\langle h \wit{h}\bigr\rangle$, because they are compatible with the Bargmann super-selection rule (\ref{gl:Barg}).
On the other hand, a co-variance requirement imposed on a {\em correlation function}, such as $C=\bigl\langle h h \bigr\rangle$, would force it
to vanish, because the Bargmann super-selection rule (\ref{gl:Barg}) cannot be satisfied.
Correlators will always be obtained by reducing them to higher response functions \cite{Pico04}.
The causality of response functions can be systematically derived from a detailed analysis of the space-time representations \cite{HenkelUnterberger,Henkel14a}.

After these preparations, we finally return to the example of the Edwards-Wilkinson model, described by the Langevin equation (\ref{gl:ew}). It is enough
to find the two- and three-point response function of the deterministic theory, from the co-variance under the generators of the
Schr\"odinger Lie algebra.\footnote{Since the deterministic part of (\ref{gl:ew}) is time-translation-invariant, the complete Schr\"odinger algebra can be used.}
First, with (\ref{gl:Rred}), the two-time response function is \cite{Hen94} (with $t>s$ because of causality \cite{HenkelUnterberger})
\beq \label{gl:Rpred}
R(t,s;\vec{r}) = r_0 \delta_{x,\wit{x}} \delta({\cal M}+\wit{\cal M})(t-s)^{-x} \exp\left[ - \frac{\cal M}{2} \frac{\vec{r}^2}{t-s}\right]
\eeq
The constraint $x=\wit{x}$ is analogous to the one following from conformal invariance.
In addition, the masses ${\cal M}=-\wit{\cal M}>0$ are related by the Bargmann rule.
These two conditions express the
relationship of the field $h$ and its conjugate response field $\wit{h}$. Second, we find the single-time correlator $C(t,\vec{r})$ with (\ref{gl:Cred}).
We need the generic three-point response \cite{Hen94} (for $\vep\to 0$ and with $t>u$ because of causality \cite{HenkelUnterberger})
\BEA
\lefteqn{\left\langle h(t+\vep,\vec{r}+\vec{r}_0) h(t,\vec{r}_0) {\wit{h}}^2(u,\vec{R})\right\rangle_0 = \delta(2{\cal M}+2\wit{\cal M})}\nonumber \\
& \times&  (t+\vep-u)^{-x} (t-u)^{-x}
\exp\left[ - \frac{\cal M}{2} \frac{(\vec{r}+\vec{r}_0-\vec{R})^2}{t+\vep-u} - \frac{\cal M}{2}\frac{(\vec{r}_0-\vec{R})^2}{t-u}\right] \nonumber \\
&\times& 
\Psi\left( \frac{\bigl((\vec{r}+\vec{r}_0-\vec{R})(t-u) - (\vec{r}_0-\vec{R})(t+\vep-u) \bigr)^2}{(t+\vep-u)(t-u) \vep}\right)
\EEA
where we already used $x_{{\wit{h}}^2}=2\wit{x}$ and $\wit{x}=x$. In the limit $\vep\to 0$
this must be finite such that the unknown scaling function $\Psi$ reduces to a constant $\Psi_0$.
We find \footnote{This corrects typos in eqs.~(30c,31) of \cite{Henkel-tutoriel}.}
\BEA
C(t,\vec{r}) &=& T\Psi_0
\int_0^t \!\D u\: u^{-2{x}} \int_{\mathbb{R}^d} \!\D\vec{R}\:
\exp\left[-\frac{\cal M}{2u}\left[\left(\frac{\vec{r}}{2}-\vec{R}\right)^2 + \left(\frac{\vec{r}}{2}+\vec{R}\right)^2 \right] \right]
\nonumber \\
&=& {T\Psi_0}\left(\frac{\pi}{\cal M}\right)^{d/2}
\int_0^t \!\D u\: u^{d/2-2{x}} \exp\left[-\frac{\cal M}{4}\frac{\vec{r}^2}{u}\right]
\nonumber \\
&=& T {c}_0\: |\vec{r}|^{d+2-4x}\, \Gamma\left( 2{x}-\frac{d}{2}-1,\frac{\cal M}{4}\frac{\vec{r}^2}{t}\right)
\label{gl:Cpred}
\EEA
where $\Psi_0$ and $c_0$ are normalisation constants and $\Gamma$ is an incomplete Gamma function.
It clearly appears that $C(t,\vec{r})$ is determined by the fluctuations in $h$ which in turn come from the noise in (\ref{gl:ew}).
But we also had to rely on consistency arguments, based on scaling,
in order to fix the unknown function $\Psi$ which is not determined by Schr\"odinger-invariance alone.

The predictions (\ref{gl:Rpred}) and (\ref{gl:Cpred}) can now be compared with the exact results of the Edwards-Wilkinson model,
readily obtained by solving (\ref{gl:ew}) \cite{Roethlein2006}. If one
identifies $x=d/2$, and matches the non-universal mass $\cal M$, the agreement is perfect \cite{Henkel-tutoriel}.
This simple example illustrates the idea how the scaling dimension $x$ of the quasi-primary
field of the Schr\"odinger group determines the functional form of the universal scaling function $F_C$ of the single-time correlator.
Two-time correlators can be treated analogously \cite{Roethlein2006}.

We close with a few further comments. \\
{\bf 1.} When quenching a magnetic system to below $T_c>0$, and the order-parameter is not conserved, the system undergoes {\em phase-ordering kinetics}, with
a dynamical exponent $z=2$ always \cite{Bray94a,Bray94b}. However, the representations of the Schr\"odinger group with generators
${\cal X}^{\rm equ}$ must be replaced by \cite{Stoimenov22-metaS,Henkel23}
\BEQ \label{gl:nrep}
{\cal X}^{\rm equ} \mapsto {\cal X} = e^{\xi \ln t} {\cal X}^{\rm equ} e^{-\xi\ln t}
\EEQ
where the generators ${\cal X}^{\rm equ}$ are those listed in table~\ref{tab:gen}. Herein,
$\xi$ serves as a further quantum number of the scaling operator these generators act on.
In this setting, one is not obliged to simply drop the time-translation generator $X_{-1}$
{}from the algebra. Rather, the breaking of time-translation-invariance occurs
`softly', since one now has
\BEQ
X_{-1}^{\rm equ} \mapsto X_{-1} = e^{\xi\ln t} \bigl( - \partial_t \bigr) e^{-\xi \ln t} = - \partial_t + \frac{\xi}{t}
\EEQ
which explicitly depends on time. This construction holds true for the entire Schr\"o\-din\-ger-Vi\-ra\-so\-ro algebra \cite{Henk06}.
In the representation (\ref{gl:nrep}), the Schr\"odinger operator also becomes time-dependent, for example
\BEQ
\mathscr{S}^{\rm equ} \mapsto \mathscr{S} = e^{\xi \ln t} \left( \partial_t - \partial_r^2\right) e^{-\xi\ln t} = \partial_t + \frac{\xi}{t} - \partial_r^2
\EEQ
This reproduces simulations in many models of phase-ordering, see \cite{HenkelPleimling2010,Henkel24}. \\
{\bf 2.} For a critical quench to $T=T_c$, in general the dynamic exponent $z\ne 2$. Since the form of the auto-response functions, determined from co-variance, only depends
on $\lambda/z$, that part of the theory can still be used, to a good degree of precision \cite{HenkelPleimling2010,HenkelPleimling01}.
However, there are indications that a better choice of representation might be a {\em logarithmic} one -- in analogy to
logarithmic conformal field theory \cite{Gurarie1993,Rahimi}, where the scaling operators become at least two-component vectors and the scaling dimensions
$x$ are replaced by Jordan matrices.
Such logarithmic representations have been constructed for the Schr\"odinger algebra\footnote{Analogous
constructions also exist for the conformal Galilean algebra, including its `exotic' central extension \cite{HenkelHosseinyRouhani2014,Hosseiny2011,Setare2011}.}
\cite{Henkel2013,HenkelRouhani2013} and indeed permit a much improved agreement with
simulation data of response functions in several critical models
($1D$ critical directed percolation \cite{Enss2004}, the $1D/2D$ Kardar-Parisi-Zhang equation \cite{HenkelNohPleimling2012,Kelling2017}
and the $2D$ critical Ising model \cite{Sastre}).

\section{Conclusions} \label{sec:conclusion}

The twin conformal and Schr\"odinger groups stand at the beginning of the systematic applications of continuous symmetry in physics, as initiated by Jacobi \cite{Jacobi} and Lie \cite{Lie}.
The pioneering work of Brinkmann \cite{Brinkmann} and of Eisenhart \cite{Eisenhart} was followed by the introduction and comprehensive use of Duval {\it et al.}'s (``Bargmann'') framework
\cite{DBKP}. This allowed, apart of finding all Schr\"odinger-symmetric mechanical systems, to study Chern-Simons vortices and fluid mechanics.
As a further example then arose the conformal Galilean group (and the recently identified meta-conformal and meta-Schr\"odinger groups), see table~\ref{tab1}.
After retracing some historical steps, and recalling several important concepts related to central extensions and super-selection rules, and whose development
took insight from quite distinct areas of physics (and mathematics)
we have seen that these three symmetries arise time and again in physical applications, only provided that there is a physical basis for emergent scale-invariance.
An important difference of Galilei- and Schr\"odinger-groups on one side and relativistic or non-relativistic conformal groups on the other, are the Bargmann super-selection rules
which can be traced back to central extensions in these non-semi-simple algebras.
Some examples, notably physical ageing, were treated  more explicitly.
Through the various applications mentioned in this review we hope to have given sufficient motivation to strive further in an ever improving
understanding and on the deep relations between them.

A major outcome of these symmetries are bootstrap approaches which try to achieve as much as possible for the symmetry itself. The most magnificent example remains of course
the conformal bootstrap in $d=2$ spatial dimensions \cite{BPZ84}, which has led to so many consequences in either $2D$ equilibrium critical phenomena or else in string theory.
It has been tried to follow these paths in different settings, notably in conformal field-theory in $d>2$ dimensions, e.g. \cite{Showk2012,Rychkov2020,RychkovSu2023},
or conformal Galilean and BMS theory, e.g. \cite{Bagchi2017,Chen2021,Bagchi:2010zz,Duval:2014uva}.

We regret to have to resist the temptation to deal with supersymmetric extensions of Galilei- and Schr\"odinger-symmetry.
A natural starting point would be the spin-$\frac{1}{2}$ L\'evy-Leblond equation \cite{LevyLeblond1967}. Since a discussion would require an article by itself, we limit ourselves to
the mere statement that these studies were initiated in
\cite{Puzalowski:1978rv,DHoker:1984vus,Beckers:1987xr,Gauntlett:1990xq,Horvathy:1992pcm,Leblanc:1992wu,Duval:1993hs,HenkelUnterberger2006}. \\[0.3cm]

After this paper was submitted, we were informed of several further researches \cite{Plyushchay-add1,Plyushchay-add2,Plyushchay-add3,Baiguera,Fedoruk,BenAchour,Dobrev} related to Schr\"odinger symmetry.
\\[0.3cm]

\noindent
{\bf Acknowledgements:}
This project was initiated jointly with Christian Duval before his untimely death.
We would like to thank Gary Gibbons for his interest, advice, and his contribution at the early stages of this work.
MH was supported by the French ANR 
UNIOPEN (ANR-22-CE30-0004-01).
PMZ was partially supported by the National Natural Science Foundation of China (Grant No. 12375084).

\newpage

\end{document}